\documentclass[aps,prb,twocolumn,longbibliography,superscriptaddress]{revtex4-2}
\usepackage{graphicx}
\usepackage{amsmath}
\usepackage{amssymb}
\usepackage{braket}
\usepackage{hyperref}
\usepackage{subfigure}
\usepackage{comment}
\usepackage{graphicx}
\usepackage{subfigure}
\usepackage[mathscr]{euscript}
\usepackage[usenames,dvipsnames]{color}
\newcommand{\asp}{}

\begin{document}

\title{Sharp detection of the onset of Floquet heating using eigenstate sensitivity}

\author{Sourav Bhattacharjee}
\email{sourav.bhattacharjee@icfo.eu}
\affiliation{ICFO - Institut de Ciències Fotòniques, The Barcelona Institute
of Science and Technology, Av. Carl Friedrich Gauss 3, 08860 Castelldefels (Barcelona), Spain}%
\author{Souvik Bandyopadhyay}
\email{sbandyop@bu.edu}
\affiliation{Department of Physics, Boston University, Boston, Massachusetts 02215, USA}%
\author{Anatoli Polkovnikov}
\email{asp28@bu.edu}
\affiliation{Department of Physics, Boston University, Boston, Massachusetts 02215, USA}%
\begin{abstract}

Chaotic Floquet systems at sufficiently low driving frequencies are known to heat up to an infinite temperature ensemble in the thermodynamic limit. However at high driving frequencies, Floquet systems remain energetically stable in a robust prethermal phase with exponentially long heating times. We propose sensitivity (susceptibility) of Floquet eigenstates against infinitesimal deformations of the drive, as a sharp and sensitive measure to detect this heating transition. It also captures various regimes (timescales) of Floquet thermalization accurately. Particularly, we find that at low frequencies near the onset of unbounded heating, Floquet eigenstates are maximally sensitive to perturbations and consequently the scaled susceptibility develops a sharp maximum. We further connect our results to the relaxation dynamics of local observables to show that near the onset of Floquet heating, the system is nonergodic with slow glassy dynamics despite being nonintegrable at all driving frequencies. 

\end{abstract}

\maketitle

\section{Introduction}\label{sec_intro}

Periodically driven quantum many-body systems are known to exhibit a plethora of interesting and novel out of equilibrium phenomena, that often have no equilibrium counterparts \cite{kitagawa_demler,takashi_review,goldman14,bukov15,eckardt17,sota19}. However, these systems inherently undergo a continuous exchange of energy with one or more external driving agents; the energetic stability of such systems, particularly in the absence of any environmental coupling, is therefore not guaranteed. For generic quantum chaotic systems, the periodic drive breaks the only local conservation law, i.e. energy conservation, thus steering the system towards a `thermal death' in the form of featureless infinite temperature state after sufficiently long times \cite{adas_heating,rigol_heating,tatsuhiko21}. Nevertheless, in finite-sized systems, { this thermal death can be avoided} if the driving frequency is sufficiently high. On the contrary, it is known that for sufficiently low frequencies, heating dynamics is 
{ effectively described by energy drift-diffusion processes, which can be captured by the Fokker-Planck equation~\cite{Bunin_2011,Hudson_2021}. As the driving frequency increases, the heating time scales rapidly increase and the Fokker-Planck approach breaks down leading to long-lived prethermal Floquet phases~ \cite{Weidinger2017,rigol_heating,bloch17,floquet_classical1,Sen_2021,prethermal_exp,adam23,he23}}. In this regime the dynamics of stroboscopically observed local observables is dictated by a local Floquet Hamiltonian \cite{machado19,bukov_prethermal,yin21}. Using perturbative arguments, the lifetime of this prethermal phase can be shown to be exponentially long in the driving frequency (see Refs.~\cite{takashi_review,Sen_2021} for review). { This lifetime can be significantly longer than experimentally relevant time-scales for sufficiently high driving frequencies such that the local Floquet Hamiltonian description becomes accurate}. At intermediate  driving frequencies, one therefore expects a transition in heating rates which separates the dynamics of the low frequency regime dictated by random unitary ensembles and the high frequency or prethermal regime dictated by a local Floquet Hamiltonian. Detection and understanding of this heating transition in Floquet systems have been the focus of numerous theoretical and experimental studies in the recent past. For example, it has been shown that the heating rate in Floquet systems might show very strong dependence
on the driving amplitudes, initial states, or resonant-type strong dependence on precise fine-tuned values of driving frequencies \cite{adas21,heyl19,time_crystal,lea21,tatsuhiko23,sengupta22}. \\

Despite considerable progress, precise detection of the driving frequencies for which typical Floquet systems start absorbing energy has remained a challenging theoretical problem. This is partly because in the low frequency regime, the dynamics starts to deviate non-perturbatively away from any local conserved Hamiltonian picture. Recent studies have therefore focused on directly observing the dynamics of local observables and their approach to infinite temperature expectations to pin-point the onset of heating. In this paper we propose a robust measure depending solely on the spectrum of the unitary propagator, to sharply detect the onset of Floquet heating in interacting periodically driven systems. Specifically, the formalism is connected to the extreme sensitivity of the Floquet unitary propagator in response to a controlled perturbation of the driving protocol near the heating transition. This is quantified using the diverging norm of the adiabatic gauge potential (AGP) or the fidelty susceptibility \cite{Campos_Venuti_2007,Gu_2008,adolfo_2012,adolfo_2017,agp1,agp2} of Floquet eigenstates near the heating transition.\\ 

For static systems, the AGP generate adiabatic transformations usually on the eigenstates of a Hamiltonian $H(\lambda)$ along some perturbation of the Hamiltonian $\partial_{\lambda}H$. Consequently, it's Hilbert Schmidt norm constitutes of the first order perturbative expansion of eigenstates with respect to $\partial_{\lambda}H$ averaged over the full spectrum. In simpler language, this norm is nothing but the typical susceptibility of energy eigenstates against an infinitesimal deformation of the Hamiltonian. It can be argued that the AGP or susceptibility is a divergent quantity in the thermodynamic limit for chaotic systems, in perfect analogy with classical chaos where phase space trajectories are exponentially sensitive to small perturbations~\footnote{ For such classical systems this susceptibility can be viewed as a complexity measure of trajectory-preserving canonical transformations~\cite{michael24}}. Consequently, it has been demonstrated in the past few years that susceptibility is a very strong and sensitive measure of the emergence of chaos in quenched quantum systems, with varying integrability breaking strengths \cite{dries_integrability,kim23}. In particular, the susceptibility reaches the maximum {inside a broad  KAM-type chaotic regime separating integrable and ergodic domains of the system. This KAM regime corresponds to a prethermal state of the system with very long relaxation times and absence of thermalization. In extensive local systems, both quantum and classical, and in the thermodynamic limit  this regime is believed to be transient. Conversely in finite-size systems this regime can be infinitely long-lived. This in turn indicates that at sufficiently weak integrability breaking, local observables fail to thermalize within the Heisenberg time even though the system can exhibit strong chaos}. \\

We consider situations in which the averaged Hamiltonian of a finite size system over a full period is chaotic; the system thus satisfies the eigenstate thermalization hypothesis (ETH) \cite{srednicki_eth,deutsch_eth,anatoli_eth,rigol_eth} in the high frequency regime when the Floquet Hamiltonian can be approximated by a finite number of terms in the Floquet-Magnus expansion. We then demonstrate how the susceptibility of eigenstates of the Floquet unitary operator sharply detect {the onset of heating transition separating two ergodic/ETH regimes corresponding to finite and infinite temperatures and described by appropriate random matrix ensembles~ \cite{rahul16,asmi23,prosen20}. { This maximum of susceptibility originates from a divergent low-frequency spectral response revealing} the existence of a robust long-lived prethermal regime with glassy relaxation dynamics near the heating transition. As in static systems, in the thermodynamic limit this prethermal KAM regime is transient but it can be stabilized by finite system sizes with the maximum of the fidelity susceptibility drifting towards higher driving frequencies as the system size increases. Mathematically the maximum of the susceptibility marks the onset of full mixing between eigenstates of the folded Floquet spectrum. Physically this corresponds to the Thouless time, setting the onset of the Random Matrix Theory (RMT) behavior of the spectrum of the Floquet unitary becoming shorter than the Heisenberg time. This Thouless time also plays the role of the relaxation time of physical observables to the infinite-temperature state. In this way there is a direct connection of the sharp detection of the heating transition with experimentally accessible dynamics of Floquet systems.} For example, local dynamics of Floquet systems have been studied recently using quantum simulators \cite{Peng2021}. We therefore believe that the method demonstrated is not only of theoretical importance, it also smoothly connects to real dynamical data within the scope of state of the art experimental setups.

{Finally, we note that the analysis and conclusions presented in this work are based on a generic model with no conservation laws, other than the emergent energy conservation at high driving frequencies.} 
\section{Methods: Adiabatic Gauge Potential in Floquet systems}\label{sec2}
 
We begin by introducing the Floquet AGP which we construct directly from the Floquet unitary propagator. This ensures that the AGP remains a meaningful quantity for all driving frequencies and even in the absence of a well-defined Floquet Hamiltonian. Without loss of generality, we consider a two-step driving protocol, where the stroboscopic evolution is governed by the Floquet  evolution operator:
\begin{equation}\label{eq_floq_unitary}
	U_F = \mathrm{e}^{-iH_B\frac{T}{2}}\mathrm{e}^{-iH_A\frac{T}{2}},
\end{equation}
where $T=2\pi/\Omega$ with $\Omega$ being the driving frequency. Let $\{\ket{\phi_n}\}$ be the set of Floquet eigenstates so that,
\begin{equation}\label{eq_floq_eigen}
	U_F\ket{\phi_n} = e^{-i\phi_n}\ket{\phi_n},
\end{equation} 
where $-\pi<\phi_1<\phi_2< \phi_3<\dots<\pi$ are the Floquet eigenphases. Here and henceforth, we shall assume natural units with both the Boltzmann and Planck constants set to unity. Note that unlike energy eigenvalues for a static Hamiltonian, the phases $\phi_n$ are not gauge invariant 
as they are defined modulo $2\pi$,
while the eigenvalues of the Floquet unitary $\exp[-i \phi_n]$ are uniquely defined. As we will see, this natural but subtle difference  is important to keep in mind while defining and analysing gauge invariant (physical) observables like the fidelity susceptibility. We now proceed to define the Floquet AGP $\mathcal A_\lambda$ as the generator of adiabatic transformations on the Floquet eigenstates along some perturbation direction $\lambda$ in the step Hamiltonians $H_A$,$H_B$ as,
\begin{equation}
    i\partial_\lambda\ket{\phi_n}=\mathcal A_\lambda\ket{\phi_n}
\end{equation}
and calculate the AGP from Eq.~\eqref{eq_floq_eigen} as,
\begin{equation}\label{eq_agp}
	\bra{\phi_m} \mathcal A_{\lambda}\ket{\phi_n}=i\bra{\phi_m}\partial_\lambda\ket{\phi_n}=i\frac{\bra{\phi_m}\partial_\lambda U_F\ket{\phi_n}}{e^{-i\phi_n}-e^{-i\phi_m}}.
\end{equation}

Without loss of generality, we assume a deformation of the half-period Hamiltonian $H_A$,
such that the Floquet unitary becomes
\begin{equation}\label{eq_perturb_H1}
	U_F(\lambda)= e^{-iH_B\frac{T}{2}}e^{-i(H_A+\lambda O)\frac{T}{2}},
\end{equation}
where $O$ is an arbitrary local operator. For example, setting $O=H_A$ we can analyze sensitivity of Floquet eigenstates to perturbatively changing the strength of the $H_A$ pulse.
It is also convenient to refer to
$\Phi_{nm}=\phi_n-\phi_m$ as the level spacing between phases of the Floquet unitary.  We reiterate that unlike the eigenvalues $e^{-i\phi_n}$ themselves the level spacing is not a gauge invariant quantity. Following simple algebraic manipulations, it is then easily seen that the matrix elements of the AGP assume the following exact gauge invariant form (see Appendix \ref{app_agp_floq} for a derivation),
\begin{equation}\label{eq_AGP_raw}
	\bra{\phi_m} \mathcal A_{\lambda} \ket{\phi_n} = -\frac{e^{i\frac{\Phi_{nm}}{2}}}{2\sin\left(\frac{\Phi_{nm}}{2}\right)}\bra{\phi_m}\mathcal{O}_A(T)\ket{\phi_n},
\end{equation}
where $\delta\lambda\,\mathcal{O}_A(T)$ represents the effective perturbation acting on the eigenstates of $U_F$ induced by the deformation $H_A\to H_A+\delta \lambda\, O$ of the driving Hamiltonian with infinitesimal $\delta \lambda$. Given $H_A\ket{E_\alpha}=E_\alpha\ket{E_\alpha}$, {the operator $\mathcal{O}_A(T)=U_F\partial_{\lambda}U_F$ can be expressed in the eigenbasis of $H_A$ as},
\begin{equation}\label{eq_O_spec}
	\mathcal{O}_A(T) = \sum_{\alpha,\beta}\bra{E_\alpha}O\ket{E_\beta}\Theta(\omega_{\alpha\beta},T)\ket{E_\alpha}\bra{E_\beta},
\end{equation}	
where $\omega_{\alpha\beta}=E_\alpha-E_\beta$ and, 
\begin{multline}\label{eq_O_mod}
	\Theta(\omega_{\alpha\beta},T) = \frac{1}{\omega_{\alpha\beta}}\Big[1-e^{i\omega_{\alpha\beta}T/2}\\+\sum_{z\in \mathbb{Z}}\left(e^{i\omega_{\alpha\beta}T/2}-i\omega_{\alpha\beta}\frac{T}{2}-1\right)\delta(\omega_{\alpha\beta}T-2z\pi)\Big],
\end{multline}
with $Z$ being an arbitrary integer. We note that the expression for $O_A(T)$ significantly simplifies if $[O,H_A]=0$. In this case it is easy to see that $\partial_\lambda U_F=-i (T/2)U_F O$ and the matrix elements of the AGP simplify to,
\begin{equation}\label{eq:simple_agp}
    \bra{\phi_m} \mathcal A_{\lambda} \ket{\phi_n} = -iT\frac{e^{i\frac{\Phi_{nm}}{2}}}{2\sin\left(\frac{\Phi_{nm}}{2}\right)}\bra{\phi_m}O\ket{\phi_n}.
\end{equation}
In this work we focus on a more generic situation for which $[O,H_A]\neq 0$.\\

The Floquet AGP defined in Eq.~\eqref{eq_AGP_raw}, in a physical sense, determines the sensitivity of the Floquet eigenstates to an infinitesimal deformation of the Hamiltonian from $H_A$ to $H_A+\delta \lambda\, O_A$. It is heavily dominated by small denominators as can be seen from Eq.~\eqref{eq_AGP_raw}. For this reason, the AGP in general is a divergent not self-averaging operator in chaotic systems. In order to regularize it, one can either consider a typical AGP~\cite{dries_integrability}) or introduce a finite time cutoff~\cite{agp2,kim23}. In this paper we adopt the latter option and define the regularized AGP as,
\begin{equation}
\mathcal A_{\lambda} = \frac{i}{2}\sum_{N=0}^{\infty}e^{-\mu N}\left[\left(\partial_\lambda U_F U^\dagger_F\right)(-N) - \left(U^\dagger_F\partial_\lambda U_F\right)(N) \right],
\end{equation}
where $N$ represent stroboscopic instants of time and $\left(\cdot\right)(N) = U_F^{\dagger N}\left(\cdot\right)U_F^N.$

Physically $\mu$ introduces the small energy or long time cutoff. To avoid divergences we choose $\mu$ to be exponentially small in the system size but always parametrically larger than the mean level spacing. It is easy to check that the Floquet AGP regularized in this way has the following matrix elements:

\begin{multline}
    \bra{\phi_m} \mathcal A_{\lambda} \ket{\phi_n} =\\- i\frac{\left(1-e^{i\Phi_{nm}}\right)\left(1+e^{-\mu}\right)}{2\left(e^{i\Phi_{nm}}-e^{-\mu}\right)\left(e^{-i\Phi_{nm}}-e^{-\mu}\right)}\bra{\phi_m}\mathcal{O}_A(T)\ket{\phi_n}\\
     \approx -{2 e^{i{\Phi_{nm}\over 2}} \sin\left({\Phi_{nm}\over 2}\right)\over \mu^2+4 \sin^2\left({\Phi_{nm}\over 2}\right)}\bra{\phi_m}\mathcal{O}_A(T)\ket{\phi_n},
\end{multline}
where the last equality follows from $\mu\ll 1$.
It is straightforward to see that in the limit $\mu\to 0$ this expression reduces to the exact definition in Eq.~\eqref{eq_AGP_raw}. Unless specified otherwise, we shall henceforth refer to the regularised version defined above as the definition of the Floquet AGP.\\

Since the AGP generates adiabatic transformations of a Floquet state, it's norm reflects the sensitivity of Floquet eigenstates against perturbations. For our purpose, we shall be interested in the Floquet fidelity susceptibility (FFS) $\chi_m$ for a given eigenstate $\ket{\phi_m}$, which is nothing but the Frobenius norm of the AGP:
\begin{align}\label{eq_chim}
\chi_m &= \sum_{n\neq m} \left|\bra{\phi_m} \mathcal A_\lambda \ket{\phi_n}\right|^2 \nonumber\\
            &\approx  \sum_{n\neq m}\frac{4\sin^2 \left(\frac{\Phi_{nm}}{2}\right)}{\left(\mu^2+4\sin^2 \left(\frac{\Phi_{nm}}{2}\right)\right)^2}\left|\bra{\phi_m}\mathcal{O}_A(T)\ket{\phi_n}\right|^2.
\end{align}

In this form it is evident that $\mu$ regularizes the norm of the AGP by eliminating divergences coming from nearly degenerate eigenstates with $|2\sin (\Phi_{nm}/2)|\lesssim \mu$. As in Ref.~\cite{agp2} we set $\mu=\gamma L\Phi_{H}/\pi$, where $\Phi_{H}$ is the the Heisenberg scale, determined numerically as the mean level spacing between the the Floquet eigenvalues $\phi_n$: $\Phi_{H}\approx T\mathcal{D}^{-1}$ in the low $T$ (ETH) regime and $\Phi_{H} \approx 2\pi/\mathcal{D}$ in the high $T$ (RMT) regime, where $\mathcal{D}$ is the Hilbert space dimension of the relevant symmetry block. In practice however, as $T$ is increased, $\Phi_H$ quickly saturates to the RMT value $2\pi/\mathcal{D}$ soon after band folding starts and much before the onset of Floquet heating. As a consequence, $\mu$ remains constant across the range of $T$ for which the crossover from ETH to RMT behavior is detected by the FFS (see Appendix  ~\ref{app_mu_vs_T}). The constant $\gamma$ is chosen to be of order one to reduce finite size effects. 
As mentioned before, one can analyze the typical fidelity susceptibility instead of introducing a finite time cutoff; the former remains well behaved even at $\mu=0$~\cite{dries_integrability,dries_mbl}:
\begin{equation}
    \chi_{typ}=e^{\overline{\log\chi_m}}.
\end{equation}
As in the case of static Hamiltonians, the fidelity susceptibility in the form~\eqref{eq_chim} can be related to the spectral function of the operator $\mathcal{O}_A(T)$ \cite{agp2} (see Appendix~\ref{app_spectral_function} for a derivation):
\begin{equation}\label{eq_chi_spec}
    \chi_m=\frac{T}{2\pi} \sum_{n\neq m}\int_{0}^{2\pi/T}\frac{4\sin^2 \left(\frac{\omega T}{2}\right)}{\left(\mu^2+4\sin^2 \left(\frac{\omega T}{2}\right)\right)^2} { \mathcal{S}_m(\omega)  d\omega},
\end{equation}
{ where the spectral function $\mathcal{S}_m(\omega)$ is given by,
\begin{multline}\label{eq_spec_orig}
\mathcal{S}_m(\omega) \\
=\sum_{n\neq m}\left|\bra{\phi_m}\mathcal{O}_A(T)\ket{\phi_n}\right|^2\delta\left(\omega T-\Phi_{nm}\mod 2\pi\right).
\end{multline}
} For numerical stability, we approximate the delta function in Eq.~\eqref{eq_spec_orig} by a continuous Lorentzian filter,
\begin{equation}
\mathcal{S}_m(\omega) = \sum_{n\neq m}\left|\bra{\phi_m}\mathcal{O}_A(T)\ket{\phi_n}\right|^2\frac{\Delta}{\Delta^2 + 4\sin^2(\frac{\omega T-\Phi_{nm}}{2})}.
\end{equation}
with broadening scale $\Delta=0.1\Phi_H$. The choice of the filter does not affect our results qualitatively (see Appendix. ~\ref{app_gaussian} for a comparison of the results with a Gaussian approximation of the delta function).
In the rest of this work, we shall be interested in the FFS and spectral function averaged over the Floquet eigenspectrum, respectively defined as,
\begin{equation}\label{eq_chi_av}
    \chi = \frac{1}{\mathcal{D}}\sum_m \chi_m, 
\end{equation}
\begin{equation}
    \mathcal{S}(\omega) = \frac{1}{\mathcal{D}}\sum_m \mathcal{S}_m(\omega)
\end{equation}
In quantum language the spectral function $\mathcal{S}(\omega)$ encodes the amount of hybridization between the eigenstates separated by the phase difference $\Phi_{nm}=\omega T ~({\rm mod}~ 2\pi)$. In turn, the fidelity susceptibility $\chi_m$ is dominated by the frequencies close to the multiples of the driving frequency $\Omega=2\pi/T$: $|(\omega T -\mu) \mod 2\pi|\ll 1$. Physically the spectral function coincides with the stroboscopic Fourier transform of the connected autocorrelation function (a.k.a. noise) of the relevant local observable  (see Appendix~\ref{app_spectral_function}). It therefore encodes information about long time relaxation dynamics of this observable \cite{dries_mbl}. Because by construction $\Phi_n, \Phi_m\in[-\pi,\pi]$,  from Eq.~\eqref{eq_spec_orig} it is easy to see that the dominant contribution to $\chi_m$ arises from $\omega\approx\mu/T$ and $\omega\approx (2\pi-\mu)/T$. Then, it can be shown (see Appendix~\ref{app_spectral_function}) that when $\mu$ is greater than the spectral gap, the FFS scales as
\begin{equation}\label{eq_chi_scaling_general}
    \chi \sim {\mathcal{S}(\mu/T)\over \mu}.
\end{equation}
Note that in deriving the above relation, we have made use of the periodic properties of the spectral function, i.e., $\mathcal{S}(\mu/T)=\mathcal{S}((2\pi-\mu)/T)$.

\section{The model}\label{sec3}
For numerical analysis, we choose the Hamiltonians $H_A$ and $H_B$ for the Floquet unitary operator in Eq.\eqref{eq_floq_unitary} as follows:
\begin{subequations} 
	\begin{multline}
		H_A = J_z\sum_{i=1}^L\sigma^z_i\sigma^z_{i+1}+J_{zz}\sum_{i=1}^L\sigma^z_i\sigma^z_{i+2} \\+J_x\sum_{i=1}^L\sigma^x_i\sigma^x_{i+1}+\left(h_x+\delta h_x\right)\sum_{i=1}^L\sigma^x_i,
	\end{multline}

	\begin{multline}
		H_B = J_z\sum_{i=1}^L\sigma^z_i\sigma^z_{i+1}+J_{zz}\sum_{i=1}^L\sigma^z_i\sigma^z_{i+2} \\+J_x\sum_{i=1}^L\sigma^x_i\sigma^x_{i+1}+\left(h_x-\delta h_x\right)\sum_{i=1}^L\sigma^x_i,
	\end{multline}
\end{subequations}
where $\sigma^x$, $\sigma^y$ and $\sigma^z$ are the Pauli matrices with interaction and field strengths chosen as $J_z = J_{zz}= 1.0$, $J_x=0.5$, $h_x=0.71$, $\delta h_x=1.0$. The parameters are chosen such that the average Hamiltonian $H_{av}=(H_A+H_B)/2$ is ergodic and satisfies ETH,

	\begin{multline}
		H_{av} = J_z\sum_{i=1}^L\sigma^z_i\sigma^z_{i+1}+J_{zz}\sum_{i=1}^L\sigma^z_i\sigma^z_{i+2} \\+J_x\sum_{i=1}^L\sigma^x_i\sigma^x_{i+1}+h_x\sum_{i=1}^L\sigma^x_i.
	\end{multline}
We set the perturbation direction $\lambda=J_z$ such that
 \begin{equation}\label{eq_probe}
    O = \frac{\partial H_A}{\partial J_z}=\sum_{i=1}^L\sigma^z_i\sigma^z_{i+1},
    \end{equation}
and we analyze the deformation of the Hamiltonian with $J_z\to 1+\delta J_z$.

For numerical analysis, we assume periodic boundary conditions. The model is therefore symmetric under translation as well as spin-inversion ($Z_2$) operations. To avoid degeneracies due to these symmetries, we restrict ourselves to a particular momentum sector $k$ such that $k\neq 0, \pi$ as these momentum sectors have the additional parity symmetry. Furthermore, we also restrict ourselves to the even $Z_2$ sector.\\

\begin{figure*}
    \centering
    \subfigure[]{
    \includegraphics[width=0.5\linewidth]{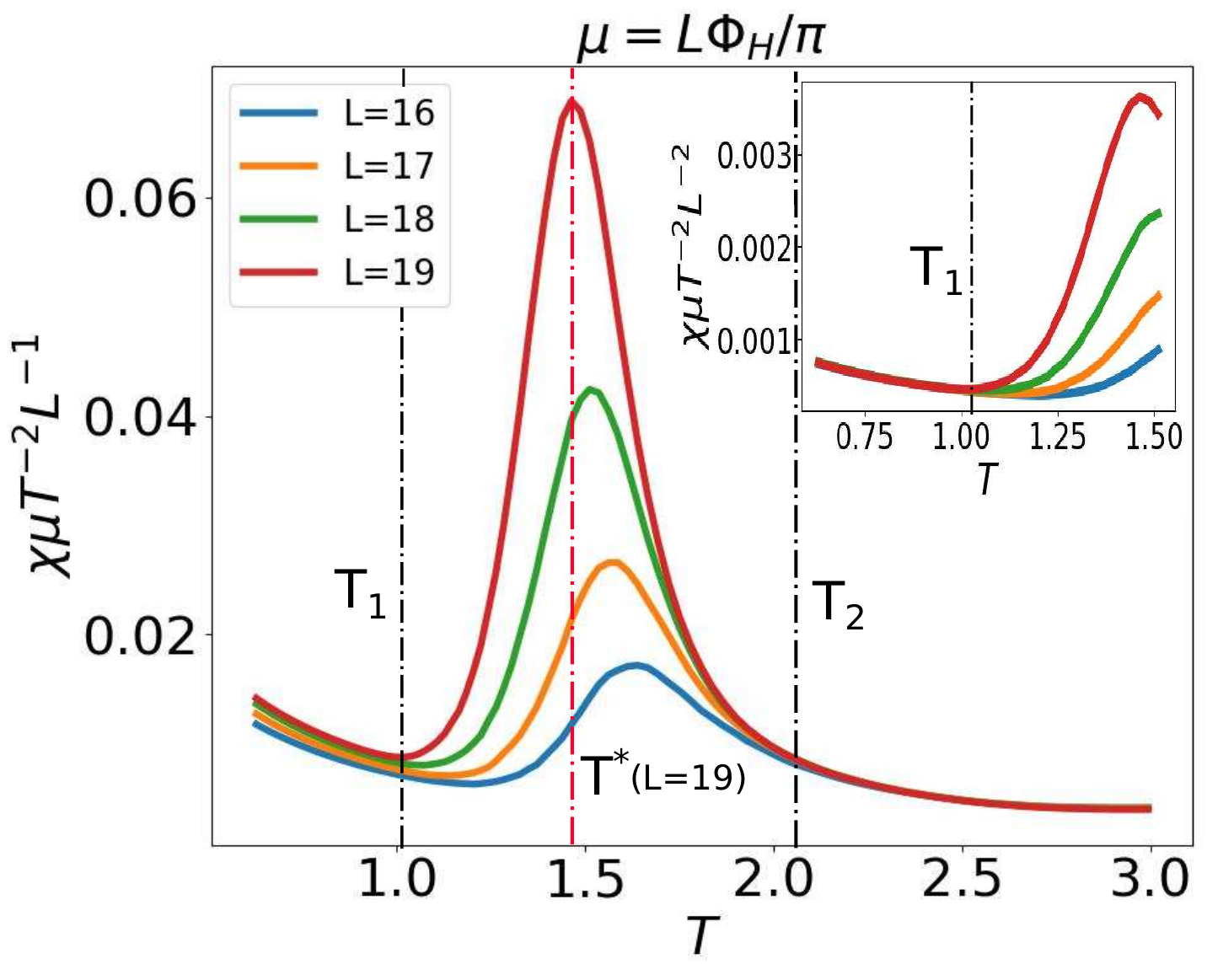}
    \label{fig_FFS_L}}%
    \subfigure[]{
    \includegraphics[width=0.5\linewidth]{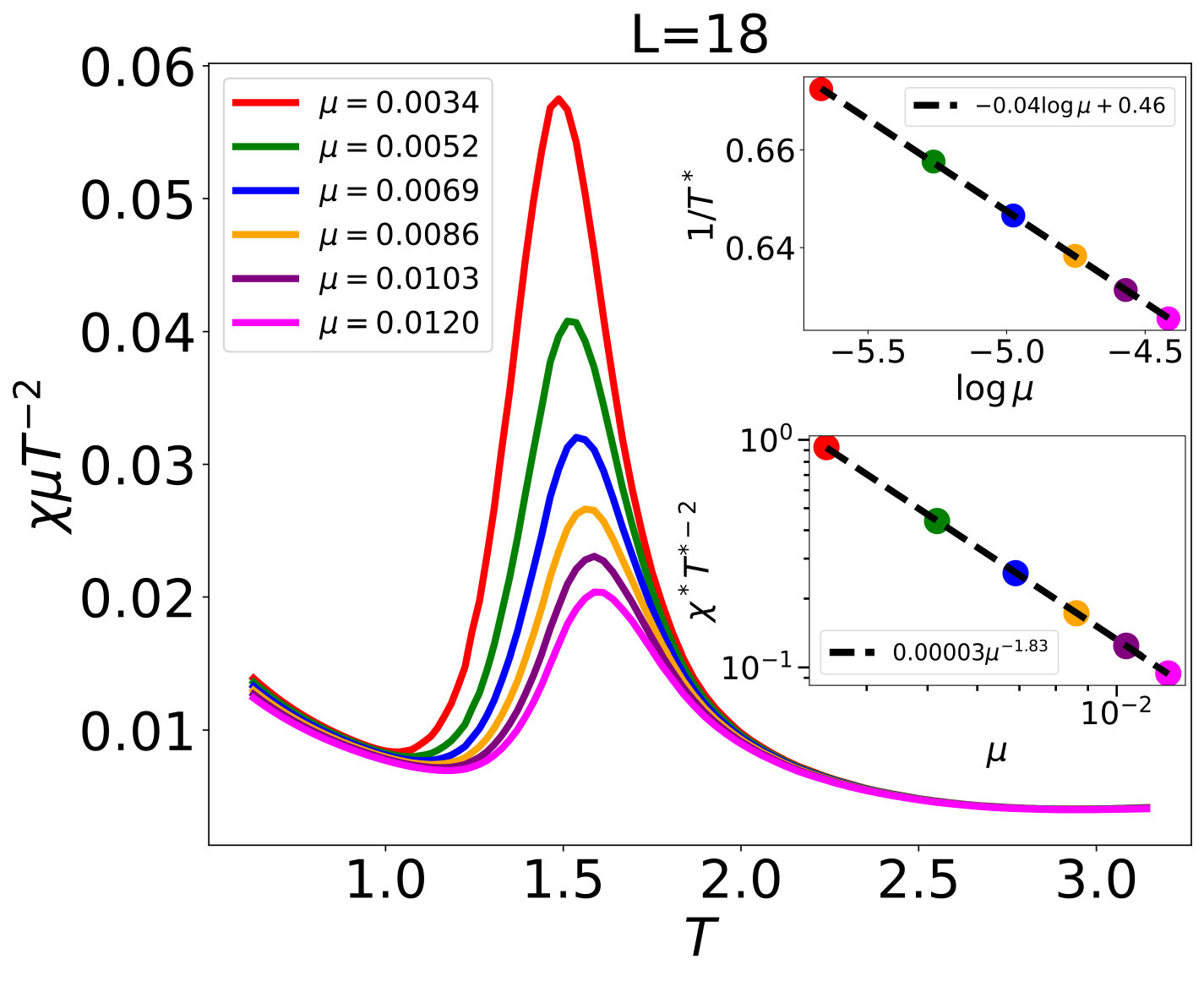}
    \label{fig_FFS_mu}}%
    \caption{The Floquet fidelity susceptibility FFS as a function of the driving period $T$ (a) for different $L$ and the cutoff fixed at $\mu=L\Phi_{H}/\pi$, and (b) for different $\mu$ with $L=18$.  The inset in (a) shows that in the high frequency regime $T<T_1\approx 1$, the FFS scales as $\chi\propto L^2/\mu$ as opposed to $\chi\sim L/\mu$ in the low frequency regime:  $T>T_2\approx 2$. The top inset in (b) shows that the position of the FFS maxima {\asp $T^\ast= 1/(\kappa_1\log(\mu^{-1})+\kappa_2)$, with $\kappa_1\approx0.04$ and $\kappa_2\approx0.46$}, drifts towards 0 with decreasing $\mu$ (as one approaches the thermodynamic limit). This implies exponentially long heating times : $\tau\propto \mu^{-1}\sim e^{1/\kappa_1T}$. The bottom inset in (b) shows that the peak of the FFS scales with the cutoff as $\chi(T^\ast)\sim \eta\mu^{-1.83}$ where $\eta\approx0.00003$.}
    \label{fig:1}
\end{figure*}

\begin{figure}
    \centering
    \includegraphics[width=\linewidth]{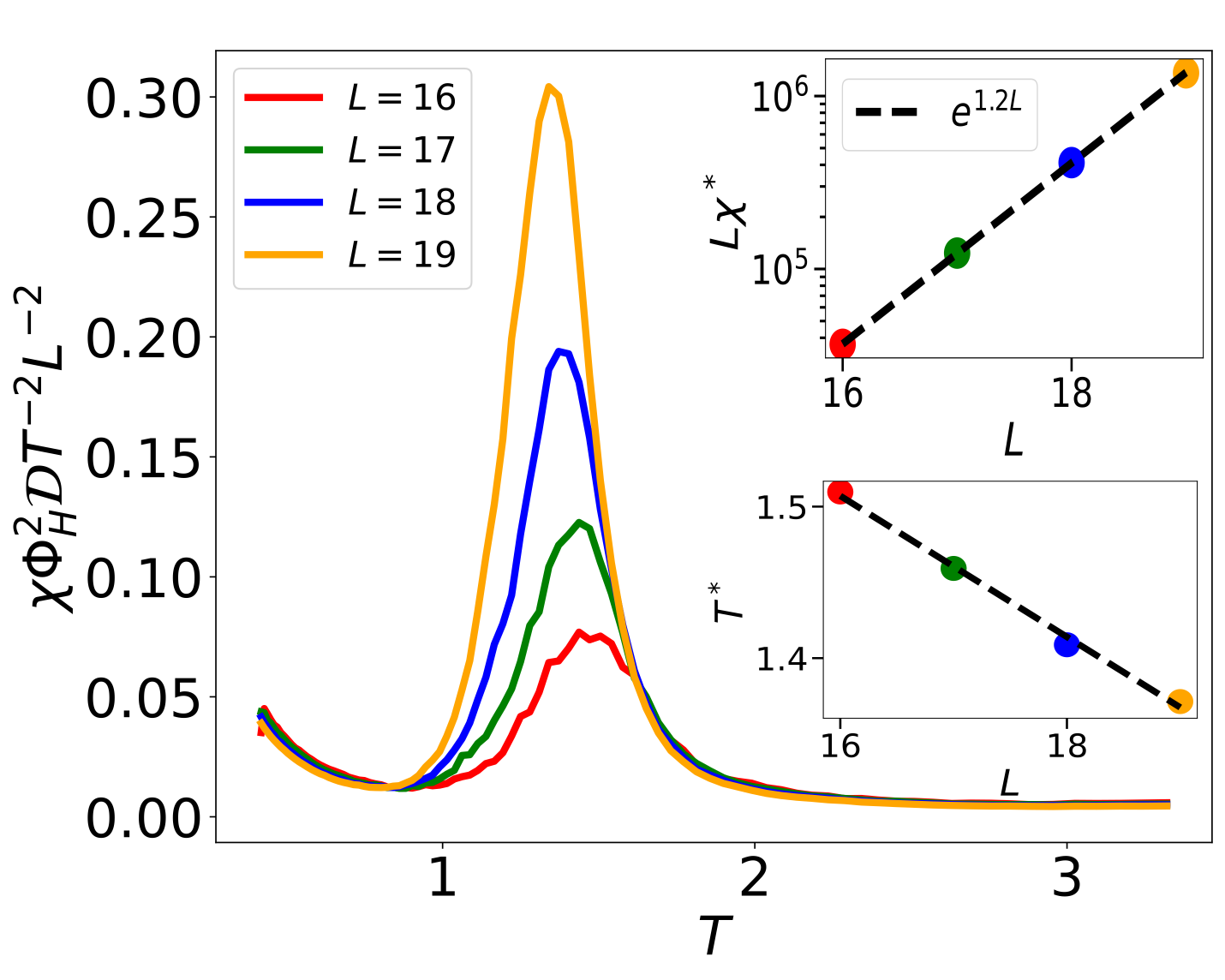}
    \caption{Typical fidelity susceptibility as function of the driving period for different system sizes. The top inset shows that the susceptibility peak scales with the system size as $e^{1.2L}/L \approx e^{2\log 2 L}/L\sim L\mathcal{D}^2$. The bottom inset shows that the position of the susceptibility peak $T^*$ decreases with increasing system size, suggesting that in the thermodynamic limit, the RMT region extends to $T\to 0$.}
    \label{fig_FFS_typical}%
\end{figure}

\section{Fidelity Susceptibility.}
In this section, we shall analyse the scaling properties of FFS for different values of the time-period $T$ and examine the corresponding heating regimes. In Fig.~\ref{fig_FFS_L} we plot the average FFS as a function of $T$ for different system sizes and the cutoff energy scale is fixed at $\mu=L\Phi_{H}/\pi$. We observe that for $T<1$, the curves for the rescaled FFS show an excellent collapse (inset of Fig.~\ref{fig_FFS_L}) suggesting that $\chi\sim L^2 T^2/\mu$. This scaling agrees with general expectations from ETH implying local thermalization and diffusive long-time relaxation of an extensive observable $O$ due to energy conservation: $\langle O(NT) O(0)\rangle\sim C \exp[-N/N_{\rm Th}]$, where $N_{\rm Th}\sim L^2/(TD)$ is the stroboscopic Thouless time and $D$ is the diffusion coefficient. As a result the low-frequency spectral function at $\omega T<\Phi_{\rm Th}\equiv 1/N_{\rm Th}\propto {\asp L^{-2}}$ {for a fixed driving period}. {\asp As it follows from Eq.~\eqref{eq:scal1} the corresponding scaling of the spectral function at $\omega T <\Phi_{\rm Th}$  is $\mathcal S(\omega)\approx C N_{\rm Th}\propto L^2$}. On the other hand, for $T>2$, the FFS is found to scale as $\chi\sim L T^2/\mu$, which now agrees with infinite temperature Floquet ETH (i.e. with a random unitary ensemble) and physically follows from the fact that in the absence of conservation laws $O=\sum_j O_j$ is a sum of local independently relaxing operators. In the intermediate range of driving periods $1<T<2$, the FFS develops a sharp peak. The position of this peak $T^\ast(\mu)$ drifts to lower periods as the cutoff $\mu$ decreases either together with the system size (panel a) or at a fixed system size (panel b): {\asp $T^\ast\sim 1/(\kappa_1\log(\mu^{-1})+\kappa_2)$, where $\kappa_1\approx0.04$ and $\kappa_2=0.46$ are non-universal constants}  (see the top inset in Fig.~\ref{fig_FFS_mu}.) 
{\asp We can invert this relation and interpret $\mu^{-1}\sim e^{1/\kappa_1T^{*}}$ as the heating time required for the system to heat up. This scaling is clearly consistent with exponential in driving frequency heating times expected in generic Floquet systems ~\cite{Abanin_2015,Mori_2016,machado19,bukov_prethermal,yin21}}.  Simultaneously the height of this peak clearly diverges faster than $1/\mu$. {Numerically this scaling is fitted best by $\chi(T^\ast)
\sim \eta/\mu^{1.83}$, with $\eta\approx0.00003$ being a non-universal constant; this scaling is close to the maximal possible divergence $\chi^*
\sim \eta/\mu^2$}~\cite{kim23,michael24}. Physically such scaling of $\chi$ signals very slow power-law or logarithmic relaxation of the system in time indicating a long-lived prethermal regime~\cite{kim23} in agreement with earlier numerical works~\cite{D_Alessio_2013,bukov_huse_asp,bukov_prethermal}.\\

Finally, in Fig.~\ref{fig_FFS_typical}, we plot the typical fidelity susceptibility for different system sizes, which also behaves qualitatively similar to the regularised FFS if we associate $\mu$ with the Heisenberg scale $\Phi_{H}/T$, demonstrating that the results are not artefacts of finite-time cutoffs. The scaling of the maximum of $\chi$ is consistent with $\chi^*\sim \exp[2\log(2) L]$ in agreement with Ref.~\cite{dries_integrability} for static systems. We shall now further elaborate on these numerically observed results in the next section.\\

\begin{figure*}
    \centering
    \subfigure[]{
    \includegraphics[width=0.5\linewidth]{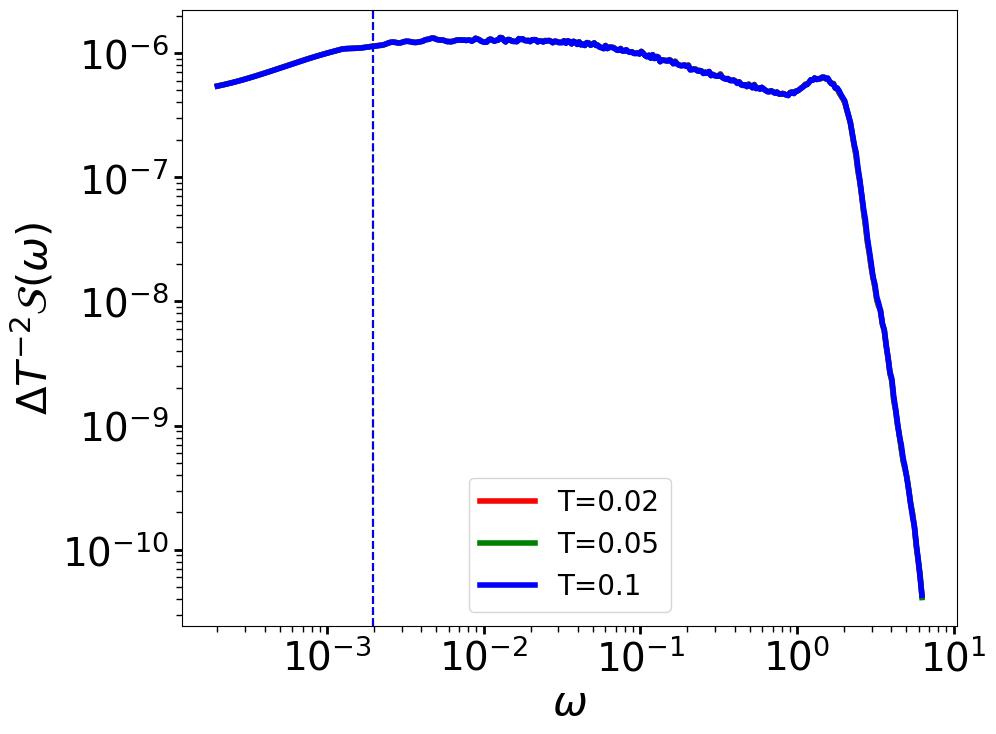}
    \label{fig_spec_high}}%
    \subfigure[]{
    \includegraphics[width=0.5\linewidth]{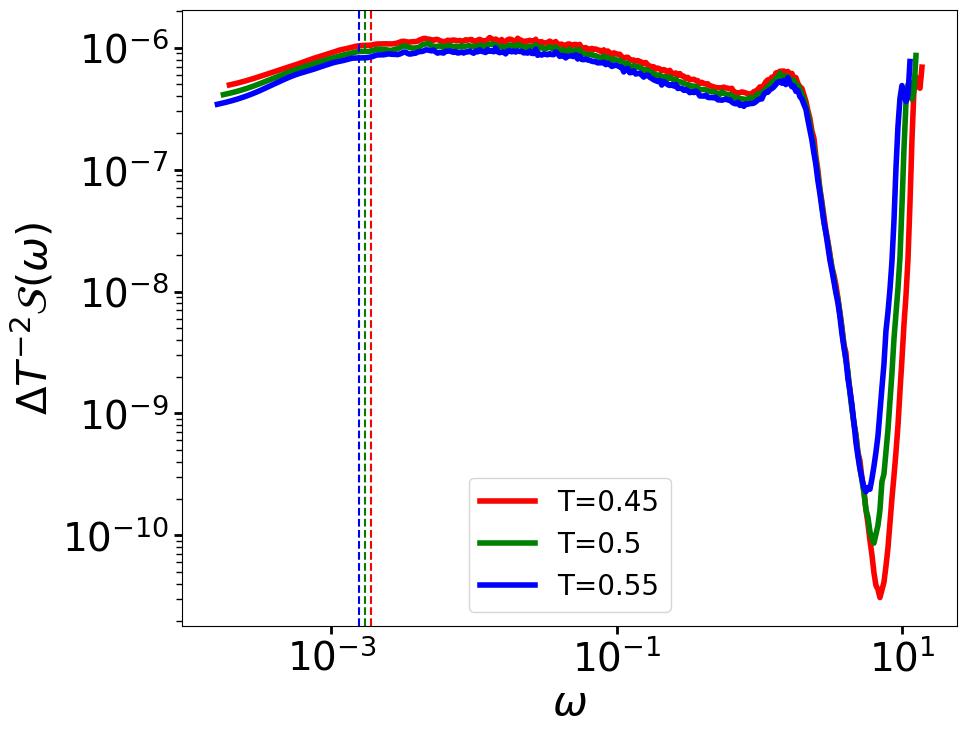}
    \label{fig_spec_mid_high_inter}}
    \subfigure[]{
    \includegraphics[width=0.5\linewidth]{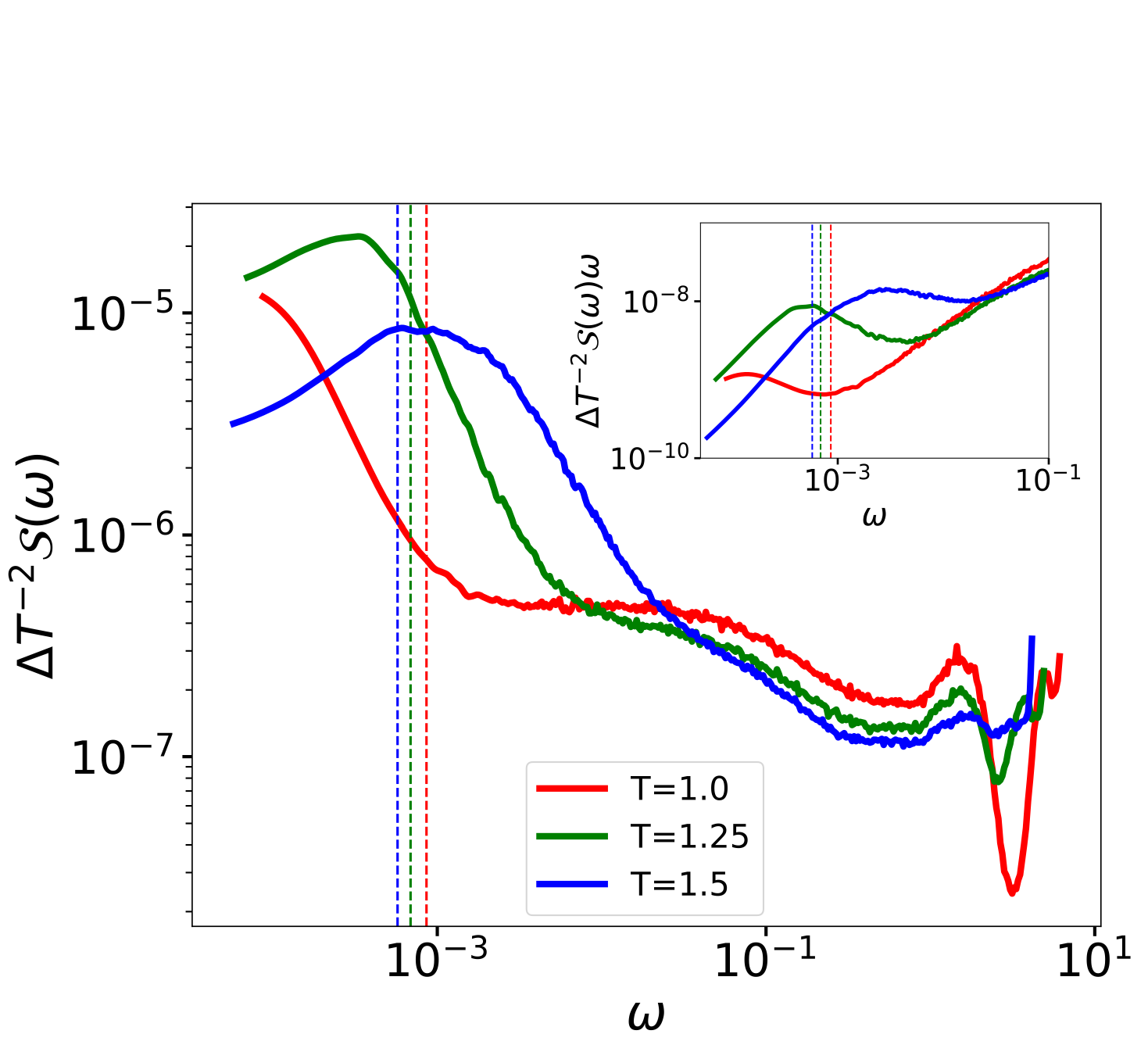}
    \label{fig_spec_inter}}%
    \subfigure[]{
    \includegraphics[width=0.5\linewidth]{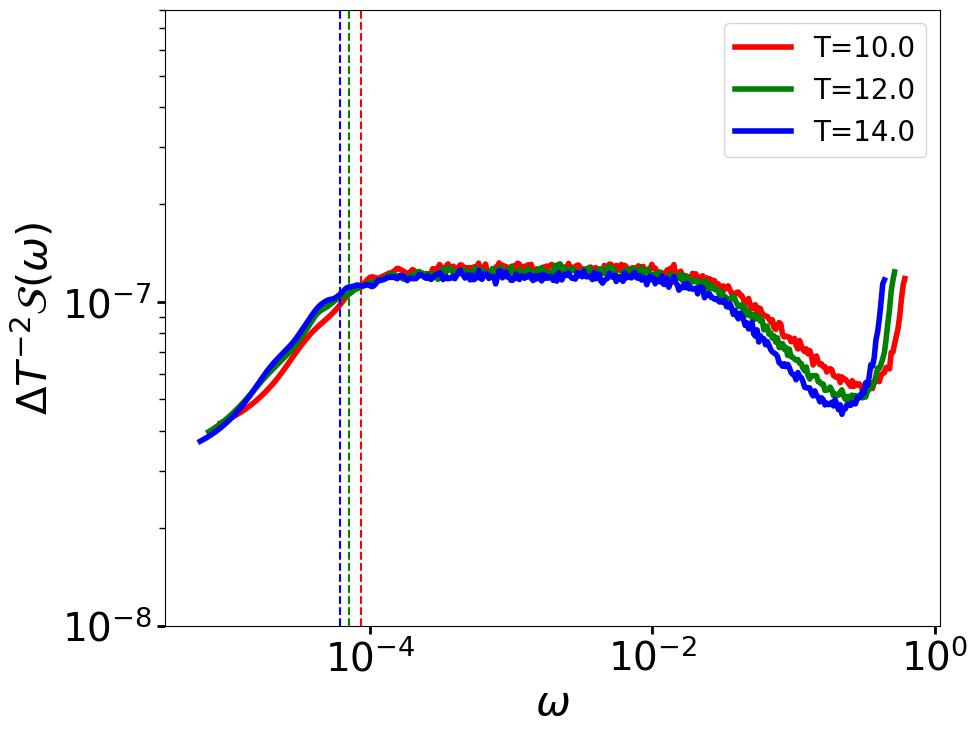}
    \label{fig_spec_low}}
    \caption{The spectral function at different driving periods for $L=18$. The dashed vertical lines mark the average level spacing (Heisenberg scale) $\Phi_{H}$ for the corresponding $T$. Note that the y-axis is scaled with $\Delta T^{-2}$ to remove the explicit scalings with $T$ arising from the matrix elements and the cutoff. For (a)$T<T'\approx 0.4 <T_1$, where $T'$ marks the beginning of band-folding, the spectral function is similar to that for quenched systems. For $T'<T<T_1\approx 1$ (b), band folding starts but spectral weight for $\omega\to 0$ remains negligible. For $T_1<T<T_2$ (c), significant spectral weight develops at low $\omega$. The inset in (c) shows the spectral function further scaled by $\omega$; the development of a flattened region (inflection) with increasing $T$ shows that as the low $\omega$ peak drifts towards $\Phi_H$ with increasing $T$, the decay of the tail of the peak approaches $1/\omega$.  Finally, for $T>T_2\approx 2$, the spectral function plateaus over the full spectrum.}
    \label{fig_spec}
\end{figure*}

\begin{figure*}
    \centering
    \subfigure[]{
    \includegraphics[width=0.45\linewidth]{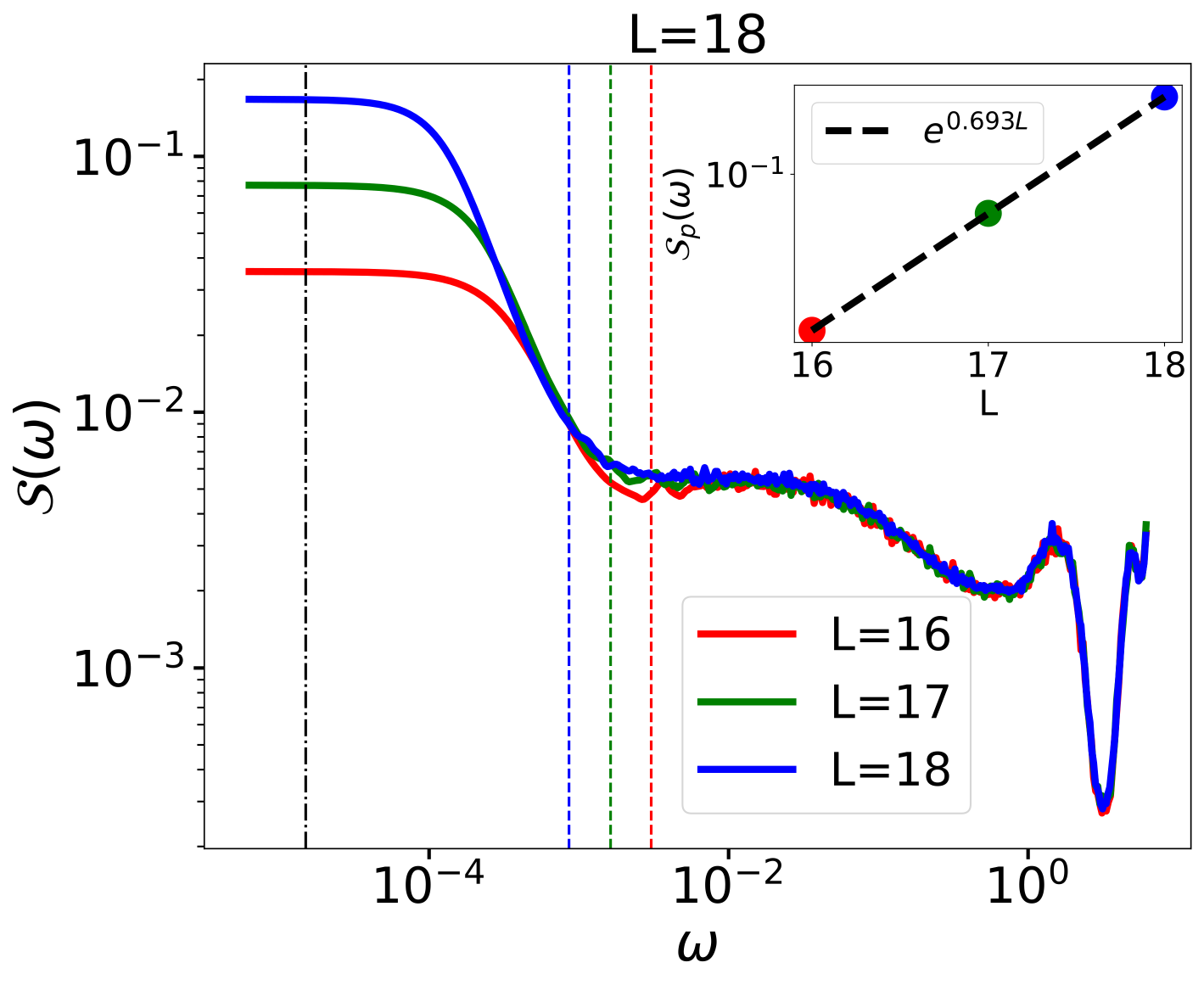}
    \label{fig_sub_heisen}}
    \subfigure[]{
    \includegraphics[width=0.45\linewidth]{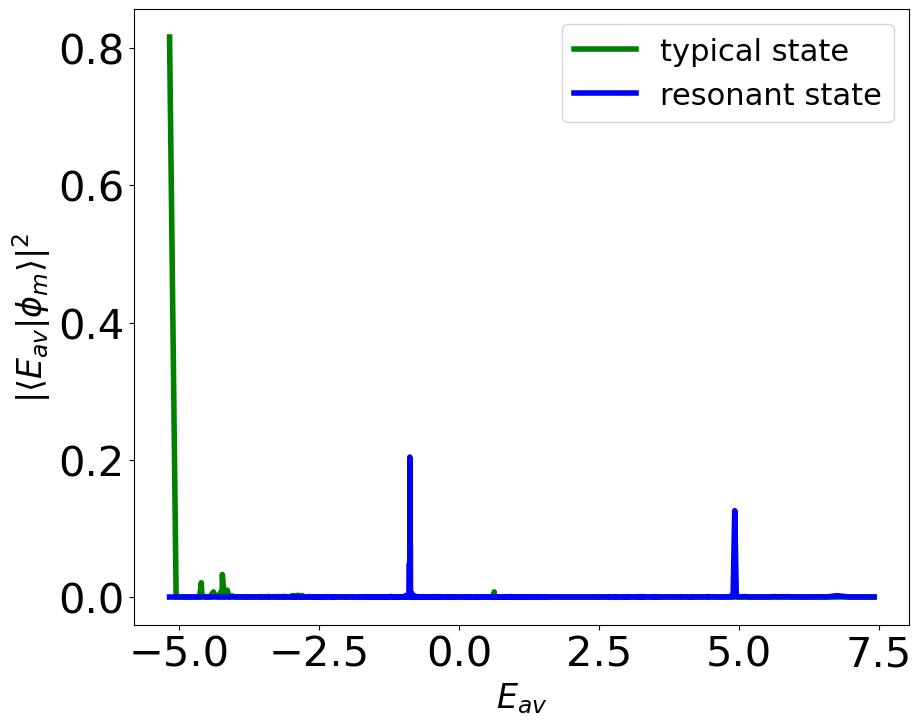}
    \label{fig_overlap}}
    \caption{(a) The spectral width at low $\omega$ for $T=1$ grows exponentially with system size. The vertical dashed lines mark the mean phase difference $\Phi_H$ for difference system sizes. The inset shows the scaling of the low $\omega$ plateau $\mathcal{S}_p(\omega)$ calculated along the vertical black dashed line as a function of the system size. (b) Overlap defined as $|\braket{E_{av}|\phi_m}|^2$ as a function of $n$ where $\ket{E_{av}}$ are the eigenstates of the average Hamiltonian and $\ket{\phi_m}$ is a Floquet eigenstate (either typical or many-body resonant) for $L=18$. The typical states (red and green) have high overlap with few eigenstates of $H_{av}$ that are close in terms of their eigenenergy. On the contrary, the many-body resonant Floquet state (blue) has overlap with distant eigenstates of $H_{av}$ having energy difference equal to $2z\pi/T$ with $z=T=1$.}
\end{figure*}

\section{Spectral function analysis} 

\subsection{High frequency driving: $T \ll 1$.}

In the high driving frequency limit, the stroboscopic evolution of the periodically modulated system is approximately described by a time-independent local Floquet Hamiltonian $H_F$ which satisfies $U_F=\exp\left(-iH_FT\right)$, and thus $H_F\ket{\phi_m}=\zeta_m \ket{\phi_m}$, where $\phi_m=T \zeta_m \mod 2\pi$. In this regime, the Floquet Hamiltonian $H_F$ can be approximated using the Floquet-Magnus expansion as \cite{takashi_review},
\begin{equation}\label{eq_HF_lowT}
    H_F = H_{av} + \sum_kT^kH_k, 
\end{equation}
where $H_{av}$ is the average Hamiltonian over one time-period and $H_k$ are higher order corrections in the expansion. It is important to note that for generic non-integrable systems, the Floquet-Magnus expansion is guaranteed to converge for $T<T' = 2\pi/W_{av}$ where $W_{av}=\max\{|\zeta_{m}^{av}-\zeta_{n}^{av}|\}\sim L$ is the bandwidth of $H_{av}$.   
In Fig.~\ref{fig_spec_high}, we indeed see that the spectral function shows a characteristic behavior that is expected in a generic non-integrable Hamiltonian -- a RMT plateau for $\Phi_{H}<\omega T <\Phi_{Th}$~\cite{anatoli_eth,serbyn21,dag23,dries_mbl}, where $\Phi_{Th}\sim T D/L^2$ is the Thouless scale. This plateau scales as $\mathcal{S}(\omega)\sim T^{2}L^2$, where quadratic dependence on $L$ reflects diffusion, while scaling with $T^2$ reflects that the effective perturbation $\mathcal O_A(T)\propto T$ [see Eq.~\eqref{eq_AGP_raw}]. This plateau leads to the ETH scaling of the fidelity susceptibility at $T\ll 1$: $\chi\sim L^2T^2/\mu$ discussed earlier (see Fig.~\ref{fig_FFS_L}). {However, as we shall demonstrate in \ref{subsec_inter}, the high-frequency ETH scaling continues to hold well beyond $T>T'$, suggesting that the dynamics is dictated by some local Hamiltonian. However, it is not apparent whether this local Hamiltonian can be obtained from the Floquet-Magnus expansion. We note in passing that a similar observation was made in Ref.~\cite{rigol_heating} in terms of level-statistics.} 
 
\subsection{Low-frequency driving: $T\gg 1$}
Although it is formally possible to identify a Floquet Hamiltonian as $H_F=i\log U_F/T$ even in this regime, the operator $H_F$ is ill defined as it is not local and does not smoothly depend on the system size and the coupling constants. The Floquet eigenphases in this regime satisfy the properties of a random circular ensemble with an average level spacing of $2\pi/\mathcal{D}$. Consequently, {the RMT plateau in the spectral function extends all the way to the full spectral bandwidth which saturates at $\omega T=2\pi$}, as can be seen in Fig.~\ref{fig_spec_low}. In other words, in the thermodynamic limit random matrix behavior is expected to hold in the full spectrum of the Floquet eigenphases. Therefore the spectral function is expected to become flat for all values of $\omega\in [0, 2\pi/T]$. Also because of lack of conservation laws the spectral function is expected to have a trivial linear scaling with $L$ due to extensivity of the perturbation.

\subsection{Intermediate frequency driving: $T\simeq 1$}\label{subsec_inter}
For $T>T'\approx 0.4$, the Floquet eigenstates start to fold and convergence of the Floquet Magnus expansion is no longer guaranteed. The folded states {may come arbitrarily close to unfolded states within the spectrum but do not immediately develop any level-repulsion with them}. This is evident from Fig.~\ref{fig_spec_mid_high_inter} { where we see that qualitative features of the spectral function remain unchanged, in particular, there is still a low frequency RMT plateau. At the same time the $T^2$ scaling of $\mathcal{S}(\omega)$ breaks down indicating that higher order terms in the Magnus expansion start affecting the Floquet Hamiltonian. { For similar reasons  the form of the operator $\mathcal O_A(T)$ also becomes affected by higher order terms.} Consequently the matrix elements $\braket{\phi_m|\mathcal O_A(T)|\phi_n}$} { entering the spectral function start depending on the driving period in a nontrivial way}. This situation continues up to $T=T_1\approx 1$, beyond which level-repulsion stars to develop between the folded states as a result of hybridization, thereby developing a low-frequency rapidly increasing tail as opposed to the ETH plateau as seen in Fig.\ref{fig_spec_inter}. This tail indicates breakdown of the Floquet-Magnus expansion and lack of existence of any local Floquet Hamiltonian, an observation which has also been reported before from the level-statistics analysis of Floquet systems \cite{rigol_heating}. This tail is also inconsistent with thermalization to the infinite-temperature phase occurring at lower driving frequencies. This spectral function analysis thus suggests that the Floquet Magnus expansion remains valid for $T \lesssim T_1$. It is interesting to note that in a previous numerical study \cite{rigol_heating}, it was observed that $T_1\approx\pi/\sigma$,  where the variance $\sigma_{av}\sim \sqrt{L}$ of the average Hamiltonian spectrum. Our observations agree with this estimate within numerical accuracy.\\

The low energy peak in the spectral function for $T>T_1$ is rather remarkable as it initially (as $T$ increases) develops much below the Heisenberg scale. The sub-Heisenberg peak is not a result of finite size effects, as can be seen from Fig.~\ref{fig_sub_heisen}. Its height scales exponentially with the system size. It appears because of Floquet many body resonances~\cite{bukov_huse_asp}, which are analogous to resonances between symmetry sectors in integrable models at small integrability breaking~\cite{Garratt_2021}. To elaborate, we first note that even for $T>T_1$, majority of the Floquet eigenstates are still adiabatically connected to those of the average Hamiltonian $H_{av}$. This can be seen from Fig.~\ref{fig_overlap}, where we show the overlap between a couple of typical eigenstates of the Floquet spectrum and that of the average Hamiltonian. {However, some of the Floquet eigenstates become hybridized mixtures of eigenstates of $H_{av}$ having energy difference equal to integer multiples of $2\pi/T$, resulting in the breakdown of adiabaticity \cite{bukov_huse_asp}. This mixing eventually leads to O(1) off-diagonal matrix elements of local observables at higher values of $T$. The appearance of such resonant states is demonstrated} in Fig.~\ref{fig_overlap} where a many-body resonant eigenstate of the Floquet spectrum has high fidelity with eigenstates of the average Hamiltonian that are distant in terms of their eigenenergy. These eigenstates have very small difference between eigenvalues of the Floquet unitary and therefore dominate late-time dynamics\\

As $T$ increases the peak is pushed beyond the Heisenberg scale indicating breakdown of this simple picture of resonances, and yet there is no full thermalization. This is true until the low frequency plateau between the Heisenberg and the Thouless scales starts emerging in the spectral function at $T\sim T_2$ and $\mathcal{S}(\omega)$ becomes flat again for { $\Phi_H\lesssim \omega T\lesssim \Phi_{\rm Th}$} as e.g. seen in Fig.~\ref{fig_spec_low}. At the point when $\Phi_{\rm Th}$ becomes equal to $\mu$ the fidelity susceptibility develops a maximum (see Fig.~\ref{fig_FFS_mu}). For the typical fidelity susceptibility (see Fig.~\ref{fig_FFS_typical}) the maximum corresponds to $\Phi_{\rm Th}\approx \Phi_H$. We can thus identify the peak of $\chi$ with the onset of thermalization/ETH (see also Ref.~\cite{dries_integrability}).
As shown in the inset of Fig. ~\ref{fig_FFS_mu}, the scaled FFS develops a sharp peak $\chi=\chi^*$ at $T=T^*$ which scales as $\chi^*\sim 1/\mu^{1.83}$ and within numerical precision agrees with $\chi^*\sim 1/\mu^{2}$ expected from general grounds~\cite{dries_integrability}. As the system size increases the condition { $\Phi_{\rm Th}=\Phi_H$} is met at lower driving periods such that the FFS peak drifts towards $T\to 0$ (see inset of Fig.~\ref{fig_FFS_typical}). This is consistent with the expectation that in the thermodynamic limit the system reaches infinite temperature state at any driving frequency.

To further support the discussion in this section, we visualize the matrix elements of $\mathcal{O}_A(T)$ in Appendix~\ref{app_mat_intense} through an intensity plot. The different regimes of heating identified through the spectral function in Fig.\ref{fig_spec} can also be clearly seen in the intensity plot. {The different regimes of heating identified through the spectral function in Fig.\ref{fig_spec} can also be clearly seen in the intensity plot. Furthermore, in Appendix~\ref{app_real_time}, we show the manifestation of these different regimes in real time stroboscopic dynamics by analyzing the connected auto-correlation function obtained from the discrete Fourier transform of the spectral function.}  

\section{Summary and concluding comments}\label{sec_discuss}

We probe the onset of non-perturbative heating in Floquet many-body systems. { We show that the heating transition can be associated with the maximum of the Floquet fidelity susceptibility (FFS), which defines sensitivity of eigenstates of the Floquet unitary to infinitesimal deformations of the driving protocol. We do the analysis by expressing the FFS through stroboscopic non-equal time spectral function of the appropriate observable, analogous to static quenches. The maximum of FFS can be either studied as a function of the time cutoff $1/\mu$ or of the system size. In the latter case it effectively corresponds to setting the time cutoff at the Heisenberg scale. In the former case one can effectively work in the thermodynamic limit. By analyzing how  the maximum of FFS with respect to the diriving period $T$ drifts with $\mu$, one can accurately identify dependence of the heating time on $T$. We also find that scaling of the FFS with the system size allows one to unambiguously identify existence of the extra conservation law at low $T$ and hence establish existence of a local Floquet Hamiltonian. }\\

{ Our analysis allows one to sharply separate three different regimes of Floquet dynamics (see Fig.~\ref{fig:1}): (i) high frequency regime with the period of the driving $T<T_1$, where  dynamics is well characterized by a local Floquet Hamiltonian and the spectrum is decsribed by the Gaussian Orthogonal Ensemble (GOE). {In this regime, diffusion washes away all temporal correlations and the connected auto-correlation vanishes beyond Thouless time. In the spectral function $S(\omega)$, this manifests as $S(\omega)\propto L^2$ and independent of $\omega$ for $\Phi_H<\omega T<\Phi_{Th}$} (ii) low frequency regime $T>T_2$ where the system heats up to an infinite temperature and the Floquet spectrum is well described by the circular unitary ensemble (CUE). {The dynamics in this regime is effectively random and again washes away all correlations, once again leading to a similar independence with respect to $\omega$, i.e., $S(\omega)\propto L$} (iii) intermediate frequency regime $T_1<T<T_2$ where the system is characterized by very long relaxation times and absence of thermalization to either of the ensembles within Heisenberg time. {In this regime, we have $\Phi_H \approx \Phi_{Th}$ and $S(\omega)\approx 1/\omega$ for $\omega T\approx \Phi_H$, suggesting development of order $\mathcal{O}(1)$ matrix elements of the local observable $O$ near the Heisenberg energy scale. This implies that the auto-correlations survive till Heisenberg time and the thermalization to infinite temperature is exponentially slow in system size.} For the model we numerically analyze, both times $T_1$ and $T_2$ decrease with the inverse time cutoff $\mu$ with $T_1(\mu)$ going to zero faster than $T_2(\mu)$ (see Fig.~\ref{fig:1}), implying that the intermediate regime becomes parametrically large at long cutoff times (large system sizes).} { Existence of robust intermediate non-thermalizing regime is consistent with earlier findings that the level spacing ratio strongly deviates from Wigner-Dyson statistics towards the Poisson regime at intermediate driving frequencies~\cite{rigol_heating}. We want to emphasize this deviation starts happening in the regime where the local Floquet Hamiltonian is ill defined and there is significant hybridization between folded energy states. We showed that the onset of the intermediate  regime corresponds to Landau-Zener type resonances, i.e. emergence of  hybridization between pairs of eigenstates of a local Floquet Hamiltonian separated by driving frequency (see also Refs.~\cite{bukov_huse_asp,Garratt_2021}). As driving frequency is lowered further these resonances proliferate forming a continuum but the system remains non-ergodic until much longer periods where full level repulsion develops and the system becomes ergodic without any conservation laws.\\

Such a prethermal non-ergodic behavior is expected in static systems for perturbations near integrable points marking the transition of the system from integrable to ETH behavior. However, note that for Floquet systems which we analyze, there is no integrable regime at any driving frequencies and the intermediate (KAM-like) phase appears at the heating transition. In a way the glassy dynamics can be viewed as coming from hybridization between different symmetry blocks due to emergent time translation invariance existing at high driving frequencies due to presence of a local Floquet Hamiltonian. These symmetry blocks play a similar role to symmetry blocks in integrable systems, which are coupled by integrability breaking perturbations. We emphasize again that breakdown of the Floquet Magnus expansion~\cite{takashi_review} is associated with the emergence of the intermediate KAM regime, which is parametrically far from the regime of full thermalization. We stress that the probe quantities presented in this paper are very well connected to the actual dynamics of local observables in Floquet systems. We therefore, believe that the predictions can be experimentally verified in present day quantum simulators. Furthermore, very recently it was observed (see Ref.~\cite{tatsuhiko23}) that Floquet states lying near the ground state of the Floquet Hamiltonian are very special in a sense that they remain robust against mixing even at significantly low driving frequencies. It might be interesting to probe the robustness of these states with the fidelity susceptibility. Furthermore, all formulations presented in this paper can also be easily extended to study classical Floquet systems \cite{floquet_classical1,floquet_classical2,michael24}.

}

\acknowledgments
We reminisce with love the memory of Prof. Amit
Dutta with whom the discussions on this project
started. We acknowledge fruitful discussions with Dries Sels, Tatsuhiko N. Ikeda, Anushya Chandran, Michael Flynn, Hyeongjin Kim, Utso Bhattacharya, Arijit Kundu, Ritajit Kundu and Arpan Das. S. Bhattacharjee acknowledges support from: ERC AdG NOQIA; MCIN/AEI (PGC2018-0910.13039/501100011033, CEX2019-000910-S/10.13039/501100011033, Plan National FIDEUA PID2019-106901GB-I00, Plan National STAMEENA PID2022-139099NB-I00 project funded by MCIN/AEI/10.13039/501100011033 and by the “European Union NextGenerationEU/PRTR" (PRTR-C17.I1), FPI); QUANTERA MAQS PCI2019-111828-2); QUANTERA DYNAMITE PCI2022-132919 (QuantERA II Programme co-funded by European Union’s Horizon 2020 program under Grant Agreement No 101017733), Ministry of Economic Affairs and Digital Transformation of the Spanish Government through the QUANTUM ENIA project call – Quantum Spain project, and by the European Union through the Recovery, Transformation, and Resilience Plan – NextGenerationEU within the framework of the Digital Spain 2026 Agenda; Fundació Cellex; Fundació Mir-Puig; Generalitat de Catalunya (European Social Fund FEDER and CERCA program, AGAUR Grant No. 2021 SGR 01452, QuantumCAT \ U16-011424, co-funded by ERDF Operational Program of Catalonia 2014-2020); Barcelona Supercomputing Center MareNostrum (FI-2023-1-0013); EU Quantum Flagship (PASQuanS2.1, 101113690); EU Horizon 2020 FET-OPEN OPTOlogic (Grant No 899794); EU Horizon Europe Program (Grant Agreement 101080086 — NeQST), ICFO Internal “QuantumGaudi” project; European Union’s Horizon 2020 program under the Marie Sklodowska-Curie grant agreement No 847648; “La Caixa” Junior Leaders fellowships, La Caixa” Foundation (ID 100010434): CF/BQ/PR23/11980043. Views and opinions expressed are, however, those of the author(s) only and do not necessarily reflect those of the European Union, European Commission, European Climate, Infrastructure and Environment Executive Agency (CINEA), or any other granting authority. Neither the European Union nor any granting authority can be held responsible for them. S. Bandyopadhyay acknowledges PMRF, MHRD India and AFOSR, USA for support. A.P. acknowledges support by the the NSF Grant DMR-2103658 and the AFOSR Grant FA9550-21-1-0342. We acknowledge the use of the QuSpin package \cite{quspin1,quspin2} for exact diagonalization.\\

\appendix

\section{AGP for Floquet systems}\label{app_agp_floq}
In this appendix, we show the detailed steps for calculating the explicit form of the Floquet AGP. We recall the general form of the FLoquet AGP from the maix text,
\begin{equation}\label{eq_app_agp}
	\bra{\phi_m} \mathcal A_{\lambda}\ket{\phi_n}=i\bra{\phi_m}\partial_\lambda\ket{\phi_n}=i\frac{\bra{\phi_m}\partial_\lambda U_F\ket{\phi_n}}{e^{-i\phi_n}-e^{-i\phi_m}}.
\end{equation}
For a peturbation of the Hamiltonian $H_A$,
\begin{equation}\label{eq_app_perturb_H1}
	U_F=e^{-iH_B\frac{T}{2}}e^{-iH_A\frac{T}{2}}\to e^{-iH_B\frac{T}{2}}e^{-i(H_A+\lambda O)\frac{T}{2}},
\end{equation}
the differential in $U_F$ assumes the form,
\begin{align}
	\partial_\lambda U_F &=-ie^{-iH_B\frac{T}{2}}e^{-iH_A\frac{T}{2}}\left[\int_0^{T/2}e^{iH_At}Oe^{-iH_At}dt\right]\nonumber\\
	&=-iU_F\left[\int_0^{T/2}e^{iH_At}Oe^{-iH_At}dt\right].
\end{align}
Given the spectral decomposition of the Hamiltonian $H_A=\sum_{\alpha}E_\alpha\ket{E_\alpha}\bra{E_\alpha}$, and $O=\sum_{\alpha,\beta}O_{\alpha,\beta}\ket{E_\alpha}\bra{E_\beta}$, the numerator in Eq.~\eqref{eq_agp} is evaluated as,
\begin{multline}\label{eq_app_h1h2_inter}
\bra{\phi_m}\partial_\lambda U_F\ket{\phi_n}=-ie^{-i\phi_m}\sum_{\alpha,\beta}O_{\alpha,\beta}\braket{\phi_m|E_\alpha}\braket{E_\beta|\phi_n}\\ \times\int_0^{T/2}e^{i(E_\alpha-E_\beta)t}dt,
\end{multline}%
where $O_{\alpha,\beta}=\bra{E_\beta}O\ket{E_\alpha}$. Let us now define the effective perturbation operator as, 
\begin{equation}
	\mathcal{O}_A(T) = \sum_{\alpha,\beta}O_{\alpha,\beta}\Theta(\omega_{\alpha\beta},T)\ket{E_\alpha}\bra{E_\beta},
\end{equation}	
where $\omega_{\alpha\beta}=E_\alpha-E_\beta$ and, 
\begin{multline} 
	\Theta(\omega_{\alpha\beta},T) = \frac{1}{\omega_{\alpha\beta}}\sum_{z\in \mathbb{Z}}\Big[1-e^{i\omega_{\alpha\beta}T/2}\\+\left(e^{i\omega_{\alpha\beta}T/2}-i\omega_{\alpha\beta}\frac{T}{2}-1\right)\delta(\omega_{\alpha\beta}T-2z\pi)\Big].
\end{multline}

Substituting in Eq.~\eqref{eq_app_agp}, we find,
\begin{align}
\bra{\phi_m} \mathcal A_{\lambda}\ket{\phi_n} &= i\frac{e^{-i\phi_m}}{e^{-i\phi_n}-e^{-i\phi_m}}\bra{\phi_m}\mathcal{O}_A(T)\ket{\phi_n}\nonumber\\
&= -\frac{e^{i\frac{\Phi_{nm}}{2}}}{2\sin\left(\frac{\Phi_{nm}}{2}\right)}\bra{\phi_m}\mathcal{O}_A(T)\ket{\phi_n}.
\end{align}

\section{Dependence of cutoff $\mu$ on $T$}\label{app_mu_vs_T}
\begin{figure}[h]
    \centering
    \includegraphics[width=\columnwidth]{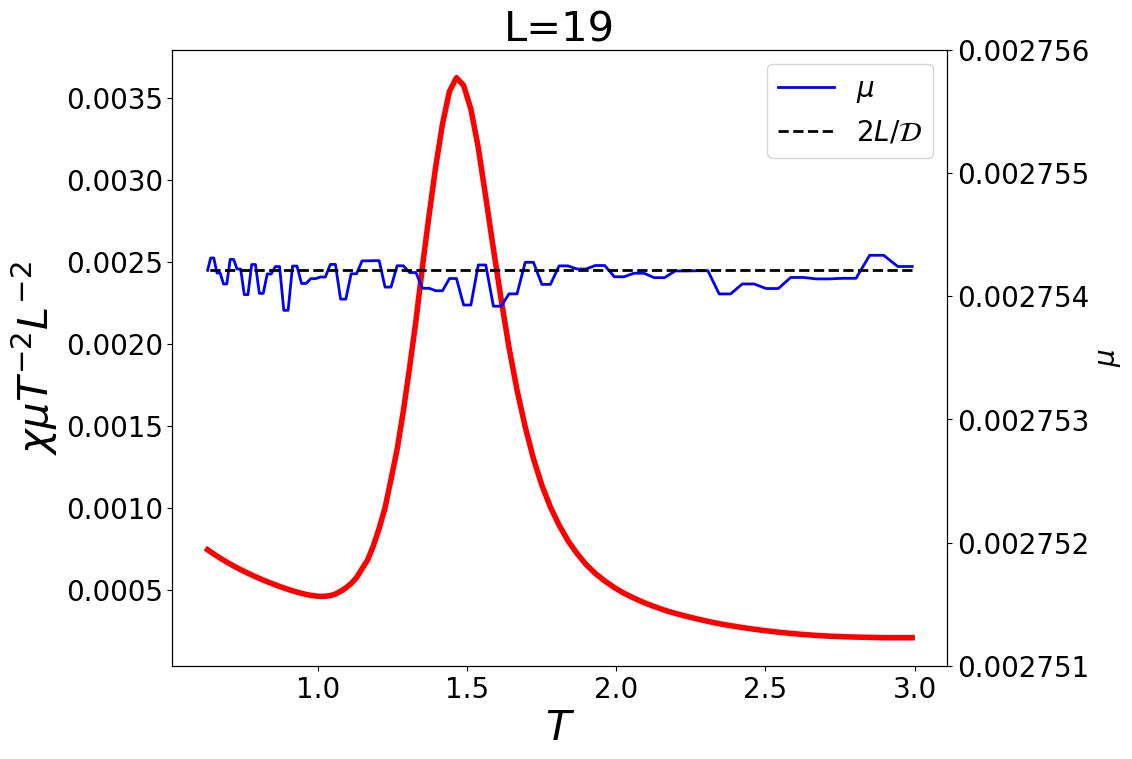}
    \caption{The finite-time cutoff $\mu=L\Phi/\pi$ remains constant throughout the parameter range of $T$ for which the FFS detects the crossover in dynamics from ETH to RMT (see red curve in Fig.~\ref{fig_FFS_L} of main text). The constant value results from the Heisenberg scale saturating to the RMT value $\Phi_H = 2\pi/\mathcal{D}$ (and thus $\mu=2L/\mathcal{D}$) for $T>T'$.}
    \label{fig_app_mu_vs_T}
\end{figure}

In this appendix, we show that the dependence of the cutoff $\mu$ with $T$ when we choose the cutoff as $\mu=\gamma L\Phi_H/\pi$. We recall that $\Phi_H\approx T/\mathcal{D}$ in the ETH regime and $\Phi_H\approx 2\pi/\mathcal{D}$ in the RMT regime. However, $\Phi_H$ quickly saturates to the RMT value as soon as $T>T'$, i.e., after band-folding starts with increasing T. As shown in Fig.~\ref{fig_app_mu_vs_T} for $\gamma=1$, the cutoff therefore remains close to $\mu\approx 2\gamma L/\mathcal{D}$ throughout the regime where the Floquet dynamics crosses over from the ETH to the RMT regime.

\begin{figure*}[t]
    \subfigure[]{
    \includegraphics[width=0.8\columnwidth]{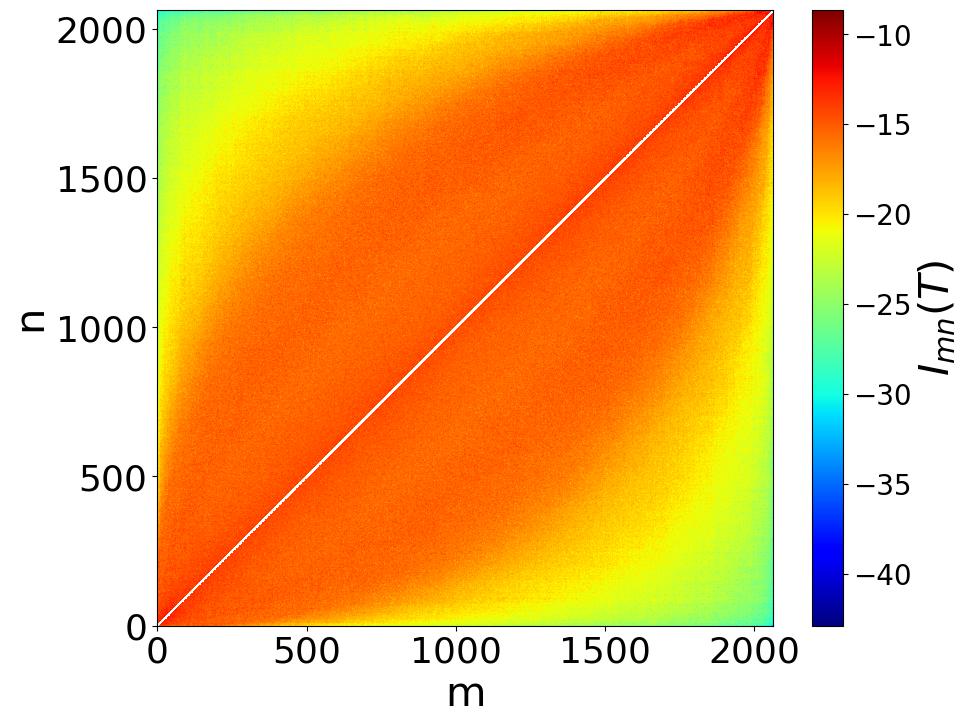}
    \label{fig_app_mat_low}}\quad\quad
    \subfigure[]{
    \includegraphics[width=0.8\columnwidth]{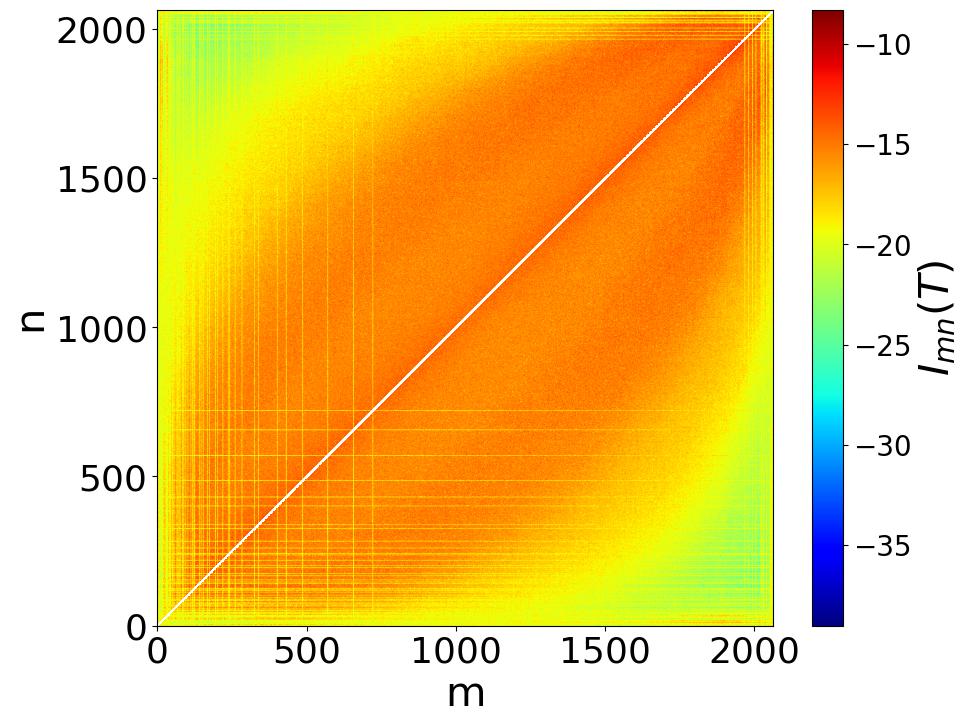}
    \label{fig_app_mat_mid}}\\
    \subfigure[]{
    \includegraphics[width=0.8\columnwidth]{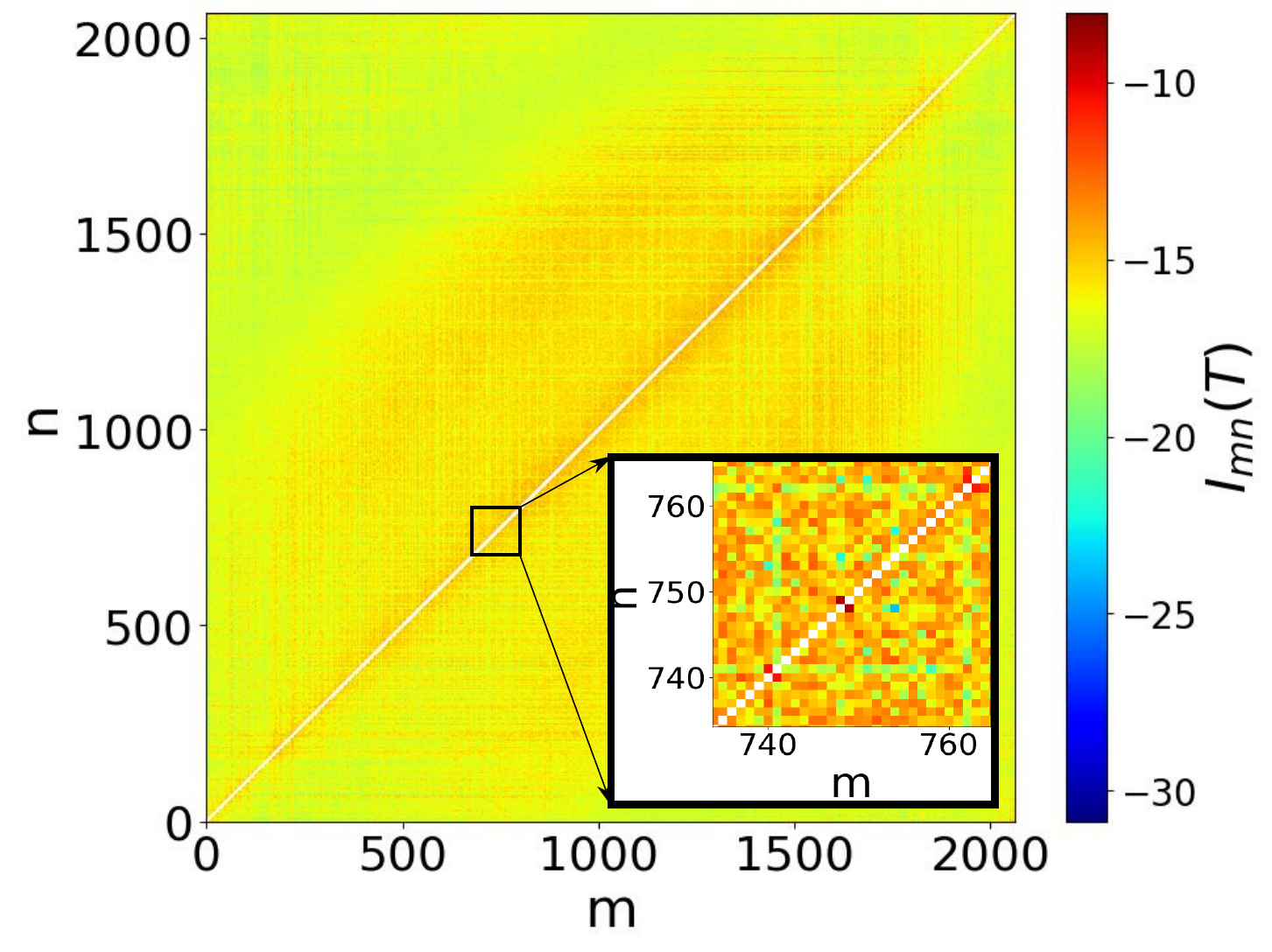}
    \label{fig_app_mat_inter}}\quad\quad
    \subfigure[]{
    \includegraphics[width=0.8\columnwidth]{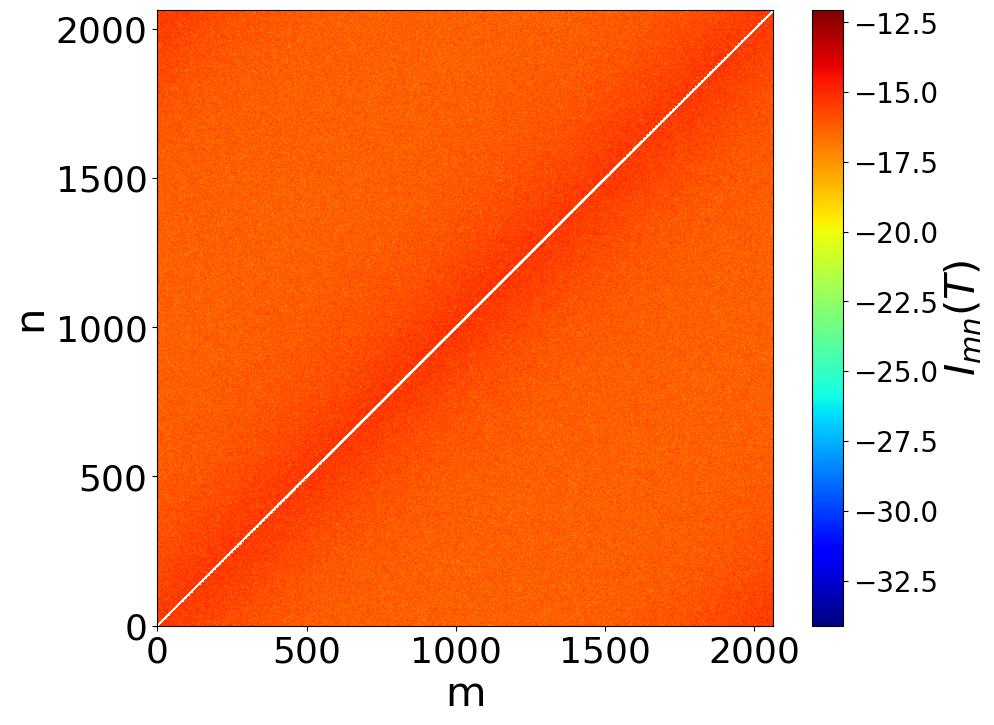}
    \label{fig_app_mat_high}}%
    \caption{Intensity plot of the matrix elements $I_{mn}(T)=\log\left[T^{-2}|\braket{\phi_m|O_A(T)|\phi_n}|^2\right]$ in the Floquet eigenbasis with $L=16$. (a) For $T=0.02$, the typical ETH profile is observed. (b) Band-folding without hybridization of the folded states can be seen for $T=0.75$ in the form of a grid-like patter. (c) Hybridization leads to many-body resonant states that have relatively higher off-diagonal elements (see bright red spots near the diagonal in the inset) at $T=1.1$. (d) Nearly uniform intensity profile for $T=5.0$ signalling the onset of RMT behavior at all scales.}
    \label{fig_app_mat_elems}
\end{figure*}

\section{Spectral function}\label{app_spectral_function}

In this appendix, we show how we obtain the spectral function from the connected auto-correlation of a local observable $O$ in an eigenstate of the Floquet unitary $\ket{\phi_n}$:
\begin{widetext}
\begin{equation}
C_o^n(N) = \frac{1}{2}\braket{\phi_n|\{O(NT)O(0)\}|\phi_n}_c=\frac{1}{2}\left[\braket{\phi_n|O(NT)O(0)+O(0)O(NT)|\phi_n}-\braket{\phi_n|O(NT)|\phi_n}\braket{\phi_n|O(0)|\phi_n}\right],
\end{equation}
\end{widetext}
where $N$ are stroboscopic counts and $T$ is the time period of the drive, such that,
\begin{equation}
O(NT) = (U^{\dagger}_F)^NO(U_F)^N.
\end{equation}

\begin{figure*}
    \centering
    \subfigure[]{
    \includegraphics[width=0.5\linewidth]{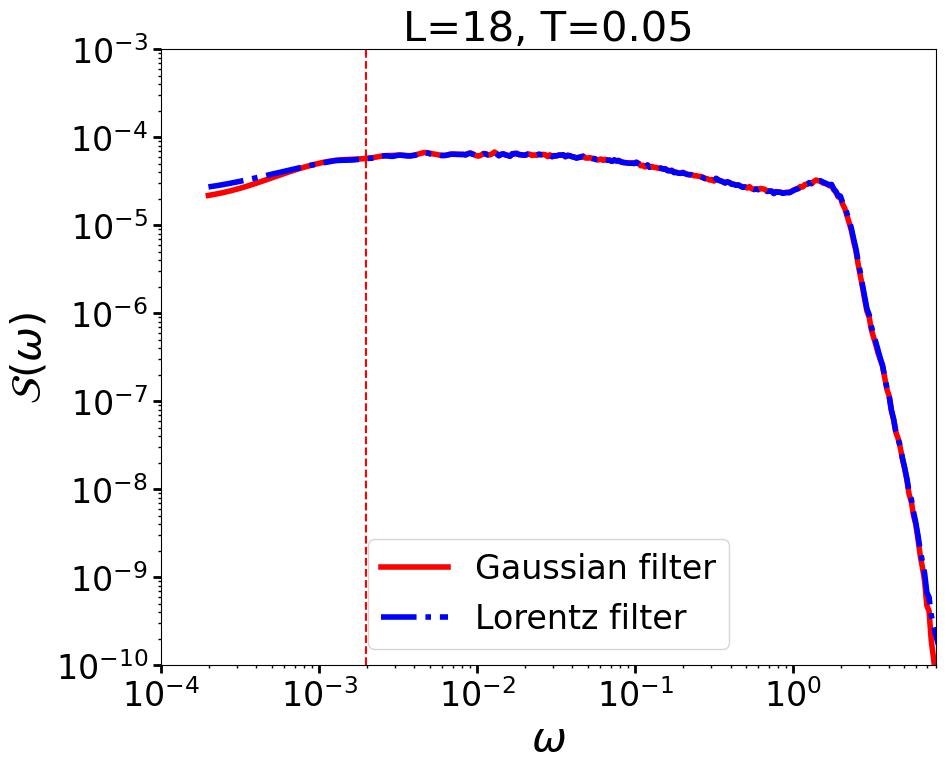}
    \label{fig_app_spec_high}}%
    \subfigure[]{
    \includegraphics[width=0.5\linewidth]{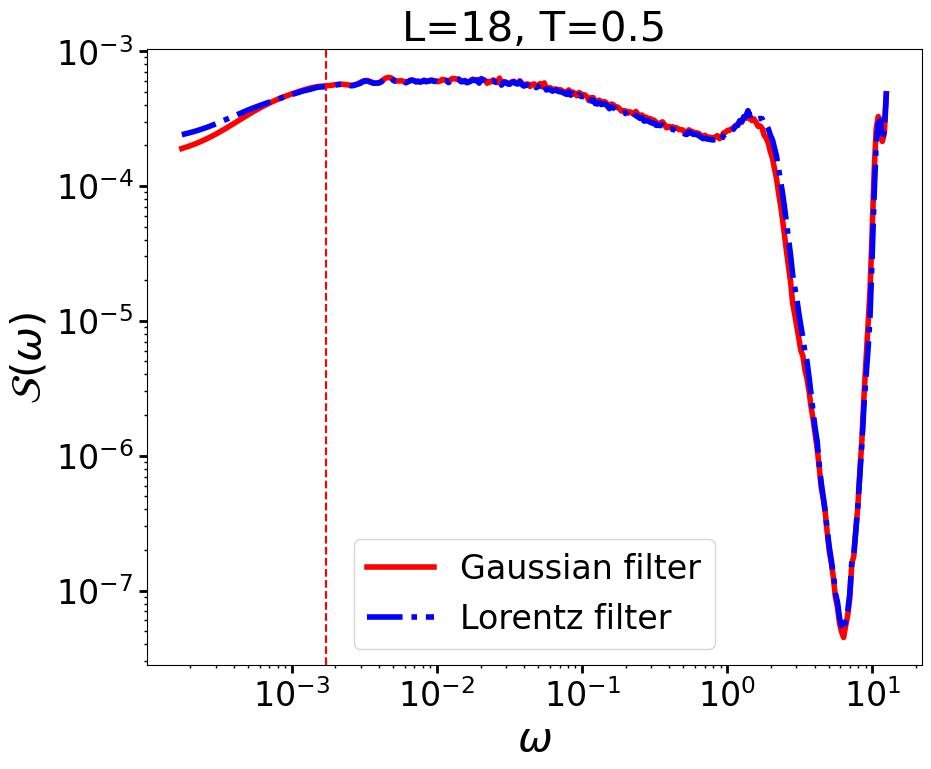}
    \label{fig_app_spec_mid_high_inter}}
    \subfigure[]{
    \includegraphics[width=0.5\linewidth]{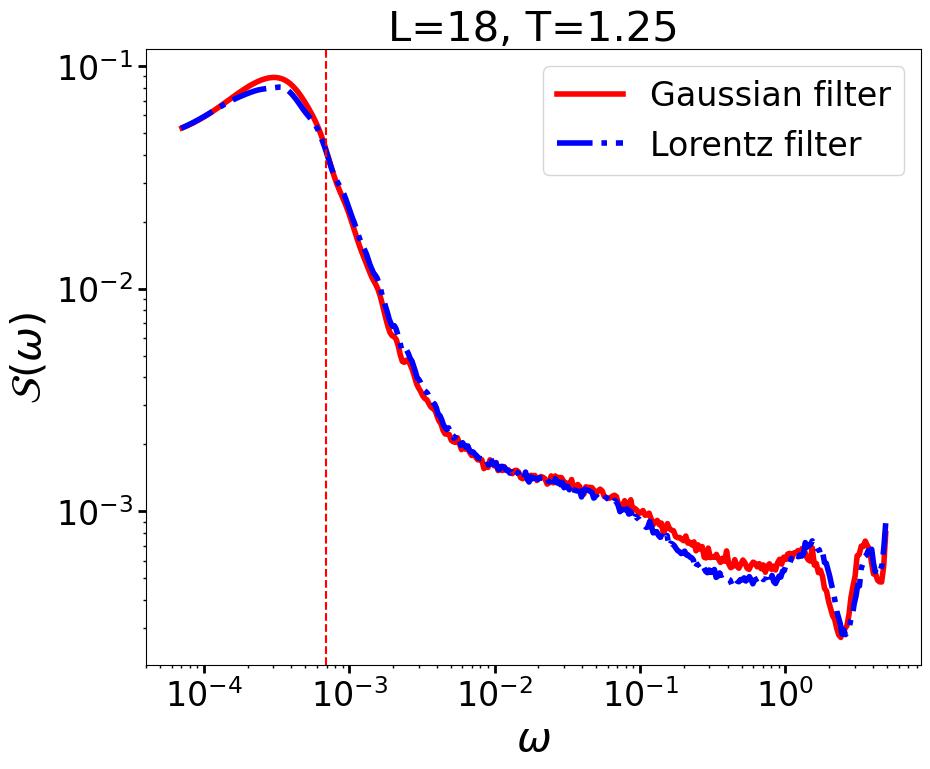}
    \label{fig_app_spec_inter}}%
    \subfigure[]{
    \includegraphics[width=0.5\linewidth]{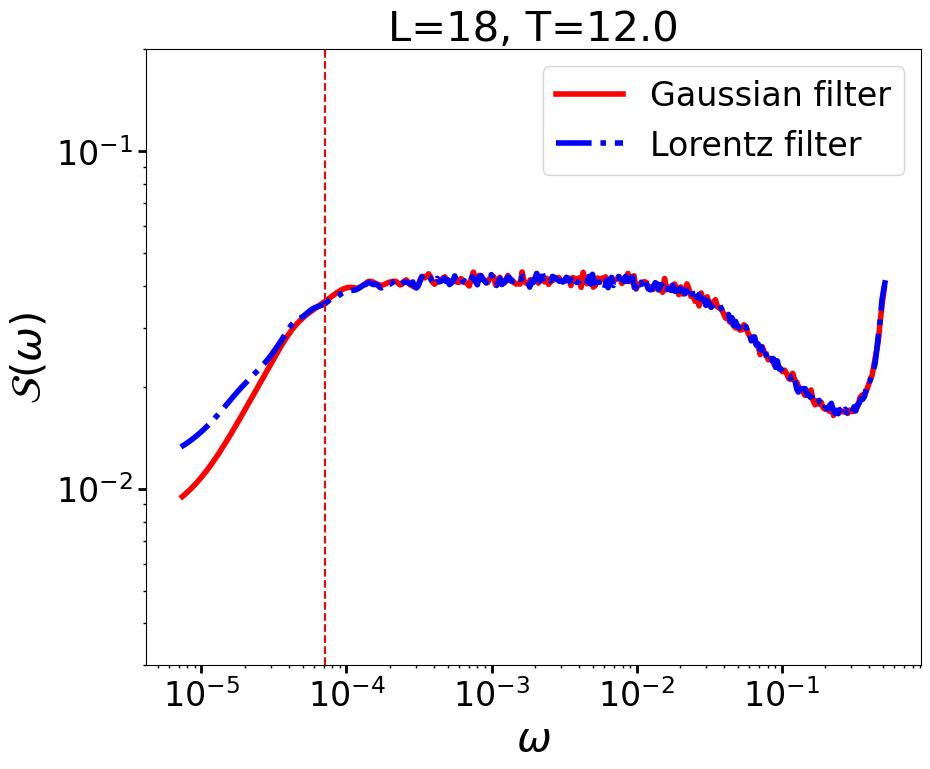}
    \label{fig_app_spec_low}}
    \caption{Comparison of the spectral function defined using a Lorentz and Gaussian filter for different values of $T$.}
    \label{fig_app_spec}
\end{figure*}

We then define the discrete Fourier transform of the the connected autocorrelations averaged over the spectrum as the spectral function:
\begin{equation}\label{eq:scal1}
\mathcal{S}(\omega) = \frac{1}{\mathcal{D}}\sum\limits_{n}\sum\limits_{N=-\infty}^{\infty}C_o^n(NT)e^{-i\omega NT}e^{-\alpha|N|T},
\end{equation}
where we introduce a broadening scale $\alpha$, which allows us to define a continuous spectral function that does not suffer from discreteness of the energy spectrum. In terms of the spectral decomposition of the Floquet unitary $U_F$, Eq.~\eqref{eq:scal1} can be simpilfied to,
\begin{equation}
\mathcal{S}(\omega) = \frac{1}{\mathcal{D}}\sum\limits_{n,m\neq m}|O_{mn}|^2\Xi,
\end{equation}
where $\Xi$ denotes the summation,
\begin{equation}
    \Xi = 2\sum\limits_{N=0}^{\infty}\cos(\Phi_{nm}N)\cos(\omega NT)e^{-\alpha NT}-1,
\end{equation}
$\Phi_{nm}$ being the phase difference between eigenvalues of $U_F$. This sum can be exactly evaluated using the cesaro summation formula and the spectral function assumes the form,
\begin{equation}
    \mathcal{S}(\omega) \sim \frac{1}{\mathcal{D}}\sum\limits_{n,m\neq m}|O_{mn}|^2\frac{\sinh(\Delta)}{\cosh(\Delta)-\cos(\Phi_{nm}-\omega T)},
\end{equation}
where we have replaced the dimensionless quantity $\alpha T$ by $\Delta$ for notational simplicity. In the limit $\Delta\rightarrow 0$, the spectral function can therefore be approximated by,
\begin{equation}
\mathcal{S}(\omega) = \frac{1}{\mathcal{D}}\sum_{m,n\neq m}\left|\bra{\phi_m}O\ket{\phi_n}\right|^2\frac{\Delta}{\Delta^2 + 4\sin^2(\frac{\omega T-\Phi_{nm}}{2})}.
\end{equation}
In the main text, we choose $\Delta=0.1\times\Phi_H$, where $\Phi_H$ is the average phase difference $\overline{\Phi_{n\, n+1}}$. Note that for $\Delta\rightarrow0$, the lorenzian weight dependent on $\Delta$ reduces to a periodic delta function and the spectral function simplifies to,
\begin{equation}\label{eq:app_sf}
\mathcal{S}(\omega) = \frac{1}{\mathcal{D}}\sum_{m,n\neq m}\left|\bra{\phi_m}O\ket{\phi_n}\right|^2\delta(\omega T-\Phi_{nm} {\text{~mod~}}2\pi).
\end{equation}
One can also see how the fidelity susceptibility $\chi_m$ of a single eigenstate as defined in the paper is connected with the spectral function for the observable $O\equiv \mathcal{O}_A$,
\begin{equation}
    \chi = \frac{1}{\mathcal{D}}\sum_{m,n\neq m}\frac{4\sin^2 \left(\frac{\Phi_{nm}}{2}\right)}{\left(\mu^2+4\sin^2 \left(\frac{\Phi_{nm}}{2}\right)\right)^2}\left|\bra{\phi_m}\mathcal{O}_A(T)\ket{\phi_n}\right|^2,
\end{equation}
which can be rewritten through the spectral function assuming $\mu\gg\Delta$ where Eq.~\eqref{eq:app_sf} applies as,
\begin{equation}\label{eq:app_chi}
    \chi = \frac{T}{2\pi}\int_{0}^{2\pi/T}\frac{\sin^2\left(\frac{\omega T}{2}\right)}{\left(\mu^2+4\sin^2 \left(\frac{\omega T}{2}\right)\right)^2}\mathcal{S}(\omega)d(\omega).
\end{equation}
In the limit $\mu\ll1$ such that we are still in the effective thermodynamic limit, the expression within the integral in the above equation \eqref{eq:app_chi} is highly peaked for small $\mu$, at frequencies such that $2\sin(\omega T/2)\sim\mu$ (i.e., for $\omega T\sim 2\sin^{-1}{\left(\frac{\mu}{2}\right)},2\pi-2\sin^{-1}{\left(\frac{\mu}{2}\right)}$ as $\omega T\in[0,2\pi]$) and it can therefore be seen that the average susceptibility scales with the time-cutoff $\mu$ for $\mu\ll1$ as,
\begin{equation}
    \chi\sim\frac{\mathcal{S}(\mu/T)}{\mu},
\end{equation}
where we have made use of the fact that $S((2\pi-\mu)/T)=S(\mu/T)$.

 \begin{figure}
    \centering
    \includegraphics[width=\linewidth]{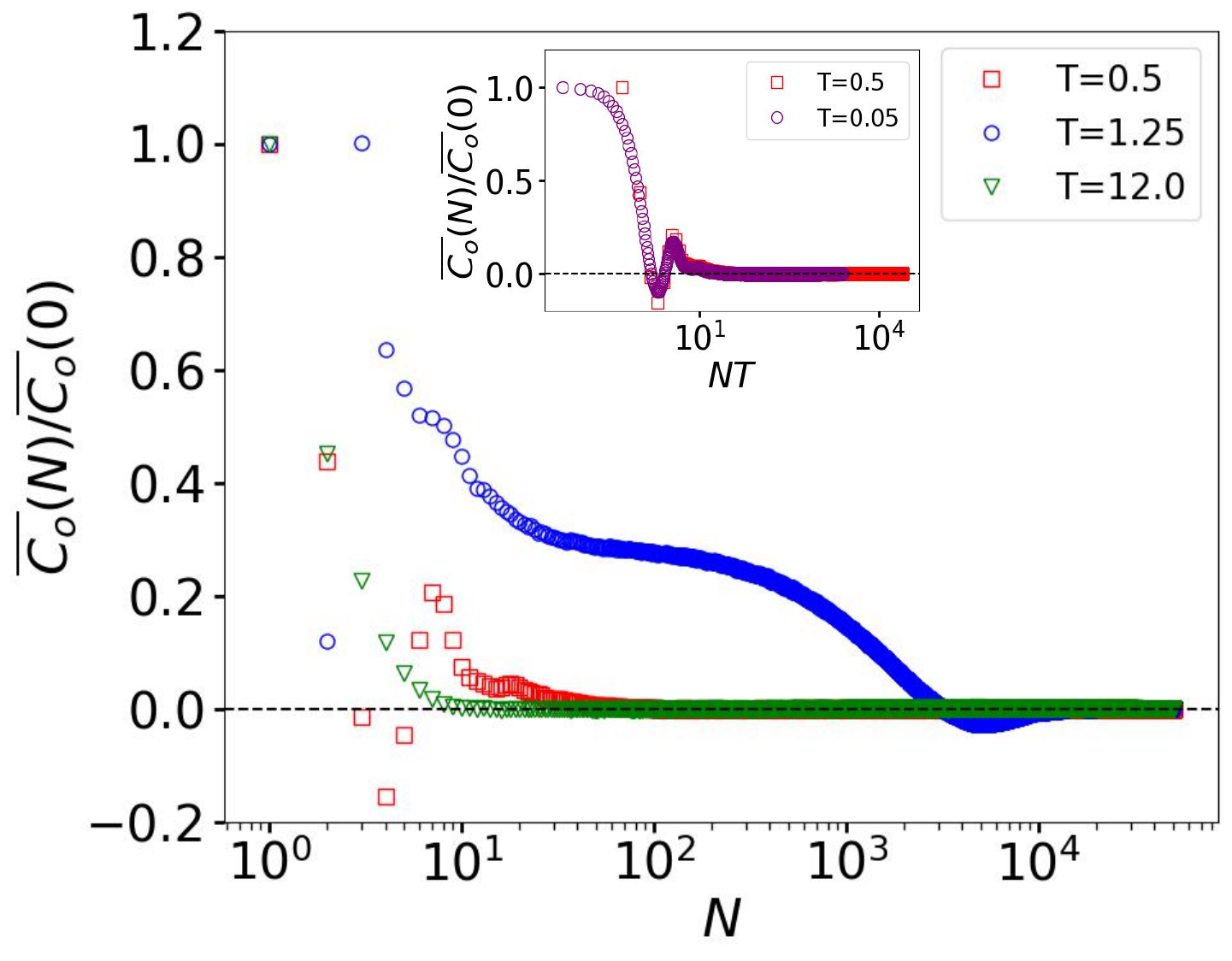}
    \caption{{Real-time stroboscopic dynamics of the connected auto-correlation function defined in Eq.~\eqref{eq_app_auto} corresponding to different regimes of heating for $L=18$. For $T=0.5$ (high driving frequency) and $T=12.0$ (low driving frequency), the connected auto-correlation function decays quickly implying fast thermalization. For $T=1.25$ (intermediate driving frequency), thermalization is slow as thouless time becomes comparable to heisenberg time (see Fig.~\ref{fig_spec}). The inset shows that below $T<T'\approx 0.4$, the evolution of the auto-correlation function with $NT$ collapses for all $T<T'$ for different driving frequencies in the high-frequency regime. This is a consequence of the fact that the heisenberg time itself depends on $T$ for $T<T'$ which leads to a trivial rescaling of the intrinsic time-scale of the dynamics.}}
    \label{fig:real_time}
\end{figure}

\section{Spectral function with gaussian filter}\label{app_gaussian}
In this appendix, we show that the results on spectral function reported in the main text does not depend upon the choice of filter used to approximate the delta function. To this end, we use a Gaussian filter in stead of the Lorentzian filter used in the main text and compare with the spectral function obtained using the Lorentz filter. To elaborate, we use a Gaussian filter of the following form for the spectral function:
\begin{equation}
    \mathcal S(\omega) = \frac{1}{D}\sum_{m,n\neq m}\left|\bra{\phi_m}O\ket{\phi_n}\right|^2\delta_g(\omega T-\Phi_{mn}),
\end{equation}
where we approximate the delta function (see Ref.~\cite{tatsuhiko23}) as,
\begin{equation} 
\delta_g(x)=\frac{1+\vartheta_3(\frac{x}{2}|\frac{i\Delta^2}{\pi})}{1+\vartheta_3(0|\frac{i\Delta^2}{\pi})}\sim\sum_{\lambda\in\mathbb{Z}}\exp\left[-\left(\frac{x-2\pi\lambda}{2\Delta}\right)^2\right],
\end{equation}
upto a normalization constant. We choose $\Delta=0.1\Phi_H$ and $\vartheta_3 (z|\tau)$ is the Jacobi theta function,
\begin{equation}
\vartheta_3 (z|\tau) = \sum_{n=-\infty}^{\infty} e^{i\pi\tau n^2}e^{2niz}.
\end{equation}

In Fig.~\ref{fig_app_spec}, we compare the plots of the spectral function using the Lorentz and Gaussian filter in all the relevant heating regimes. It is straightforward to see that the spectral functions obtained from the two different filters are almost identical and have the same qualitative behavior.

\section{Visualization of matrix elements of the perturbation operator}\label{app_mat_intense}
In Fig.~\ref{fig_app_mat_elems}, we show the intensity $I(T)$ of the matrix elements of the perturbation operator $O_A(T)$, defined as,
\begin{equation}
I_{mn}(T)=\log\left(\frac{\left|\braket{\phi_m|\mathcal{O}_A(T)|\phi_n}\right|^2}{T^2}\right),
\end{equation}
for $L=16$. For $T< T'$, the matrix elements decay similarly to that expected from ETH behavior, as shown in Fig.~\ref{fig_app_mat_low} for $T=0.02$. After band-folding begins above $T'$, the folded states do not immediately hybridize upto $T_1$ as can be seen from the appearance of grid-like patterns in Fig.~\ref{fig_app_mat_mid} with $T=0.75$. On further increasing $T$ beyond $T_1$, hybridization leads to the emergence of many-body resonant states as can be seen in the inset of Fig.~\ref{fig_app_mat_inter} (bright red spots near the diagonal) for $T=1.1$. Finally, RMT behavior emerges beyond $T>T_2$ when hybridization is complete leading to a featureless profile of the matrix elements, as shown in Fig.~\ref{fig_app_mat_high} for $T=5.0$.\\

\section{Real time stroboscopic dynamics}\label{app_real_time}
{In this appendix, we analyze the real time stroboscopic evolution of the local observable $\mathcal{O}_A(T)$ correspond to the different regimes of heating in Fig.~\ref{fig_spec} in the main text. Taking a discrete Fourier transform of Eq.~\eqref{eq:scal1}, we obtain,
\begin{equation}\label{eq_app_auto}
    \overline{C_o}(N)\equiv\frac{1}{\mathcal{D}}\sum_nC_o^n(NT)e^{-N\Delta}=\frac{1}{N_{\rm max}}\sum_{\omega}\mathcal{S}(\omega)e^{i\omega NT},
\end{equation}
where $\omega=0, 2\pi/N_{\rm max}T, 4\pi/N_{\rm max}T,\dots,2\pi/T$ with $N_{\rm max}$ being the number of discrete intervals used in the Fourier transform. In other words, the discrete Fourier transform of the spectral function provides the connected auto-correlation function of the same observable as shown in Fig.~\ref{fig_spec} with an overall damping $e^{-N\Delta}$ that is negligible for $N\Delta\ll 1$. In Fig.~\ref{fig:real_time}, we plot $\overline{C_o}(N)$ as a function of $N$ to show the evolution of the auto-correlation function in different heating regimes. The fast thermalization for $T<T_1\approx 1$ as well as for $T>T_2 \approx 2$ is clearly evident in the form of exponentially decaying auto-correlation function and saturation after finite Thouless time. However in the transition regime $T=1.25$, the decay is slower as the Thouless time-scale merges with the Heisenberg scale. We note that for $T<T'$, the bandwidth of the Floquet operator $U_F$ decreases with $T$, and consequently the Heisenberg time increases linearly with $T$ in leading order of the high-frequency expansion. This results in a trivial rescaling of the time-scale of the dynamics, which in turn results in a slower decay of auto-correlation function with decreasing $T$ as shown with the scaled collapse in the inset of Fig.~\ref{fig:real_time}}. \\

\noindent\textbf{Author contribution:} AP conceived the idea. All the authors contributed equally to the study and preparation of the paper.\\

\noindent\textbf{Data availability:} The authors declare that the findings of this study are based on numerical data that can be easily reproduced using the techniques discussed in the paper.\\

\bibliography{sn-bibliography}

\end{document}